\documentclass[twocolumn]{aastex7}

\usepackage{soul} 
\usepackage{amsmath}
\usepackage{amssymb}
\usepackage{xspace}
\usepackage{xifthen}
\usepackage{gensymb}
\usepackage{adjustbox}
\usepackage{caption}

\newcommand{\ie}{i.e.\xspace}
\newcommand{\eg}{e.g.\xspace}

\mathchardef\mhyphen="2D

\newlength{\dhatheight}

\newcommand{\unit}[1]{\ensuremath{\mathrm{\,#1}}\xspace}
\newcommand{\Myr}{\unit{Myr}}
\newcommand{\Gyr}{\unit{Gyr}}

\newcommand{\masyr}{\unit{mas}~\unit{yr}^{-1}\xspace}

\newcommand{\km}{\unit{km}}
\newcommand{\kms}{\km \second^{-1}}

\newcommand{\second}{\unit{s}}

\newcommand{\magn}{\unit{mag}}

\newcommand{\e}{\unit{e^{-}}}

\newcommand{\secref}[1]{Section~\ref{sec:#1}}

\newcommand{\bandvar}[2][]{%
  \ifthenelse{\isempty{#1}}{\var{#2}}{\var{#2\_#1}}%
}

\newcommand{\var}[1]{\ensuremath{\texttt{\MakeUppercase{#1}}}\xspace}

\providecommand\physrep{\ref@jnl{Phys.~Rep.}}%
\providecommand\apjs{\ref@jnl{ApJS}}%
\providecommand{\jcap}{\ref@jnl{JCAP}}%

\usepackage[symbol*]{footmisc}

\shorttitle{Extra-tidal features around GCs}
\shortauthors{Chiti, Tavangar et al.}

\graphicspath{{./}{figures/}}

\begin{document}

\title{DELVE-ing into the Milky Way's Globular Clusters: Assessing extra-tidal features in NGC 5897, NGC 7492, and testing detectability with deeper photometry}


\author[0000-0002-7155-679X]{A. Chiti$^*$$^\dagger$}\email{achiti@uchicago.edu}
\affil{Department of Astronomy \& Astrophysics, University of Chicago, 5640 S Ellis Avenue, Chicago, IL 60637, USA}
\affil{Kavli Institute for Cosmological Physics, University of Chicago, Chicago, IL 60637, USA}\footnotetext[1]{These authors contributed equally to this work}
\affil{Kavli Institute for Particle Astrophysics \& Cosmology, Stanford University, Stanford, CA 94305, USA}
\footnotetext[2]{Brinson Prize Fellow}

\author[0000-0001-6584-6144]{K. Tavangar$^*$}\email{k.tavangar@columbia.edu}
\affiliation{Department of Astronomy, Columbia University, 550 West 120th Street, New York, NY, 10027, USA}

\author[0000-0001-6957-1627]{P. S. Ferguson}\email{pferguso@uw.edu}
\affiliation{Department of Astronomy, University of Washington, Seattle, WA 98195, USA}

\author[0000-0002-3690-105X]{J. A. Carballo-Bello}\email{jcarballo@academicos.uta.cl}
\affiliation{Instituto de Alta Investigación, Universidad de Tarapacá, Casilla 7D, Arica, Chile}

\author[0009-0004-5519-0929]{A. M. Senkevich}\email{a.senkevich@surrey.ac.uk}
\affil{Department of Physics, University of Surrey, Guildford GU2 7XH, UK}

\author[0000-0002-8448-5505]{D. Erkal}\email{d.erkal@surrey.ac.uk}
\affil{Department of Physics, University of Surrey, Guildford GU2 7XH, UK}

\author[0000-0001-8251-933X]{A. Drlica-Wagner}\email{kadrlica@fnal.gov}
\affil{Department of Astronomy \& Astrophysics, University of Chicago, 5640 S Ellis Avenue, Chicago, IL 60637, USA}
\affil{Kavli Institute for Cosmological Physics, University of Chicago, Chicago, IL 60637, USA}
\affiliation{Fermi National Accelerator Laboratory, P.O. Box 500, Batavia, IL 60510, USA}

\author[0000-0002-6021-8760]{A. B. Pace}\email{apace@virginia.edu}
\affil{Department of Astronomy, University of Virginia, 530 McCormick Road, Charlottesville, VA 22904, USA}

\author[0000-0002-4863-8842]{A.~P.~Ji}\email{alexji@uchicago.edu}
\affiliation{Department of Astronomy \& Astrophysics, University of Chicago, 5640 S. Ellis Avenue, Chicago, IL 60637, USA}
\affiliation{Kavli Institute for Cosmological Physics, University of Chicago, 5640 S. Ellis Avenue, Chicago, IL 60637, USA}

\author[0000-0003-4102-380X]{D. J. Sand}\email{dsand@arizona.edu}
\affiliation{Steward Observatory, University of Arizona, 933 North Cherry Avenue, Tucson, AZ 85721-0065, USA}

\author[0000-0002-9269-8287]{G. Limberg}\email{limberg@uchicago.edu}
\affil{Department of Astronomy \& Astrophysics, University of Chicago, 5640 S Ellis Avenue, Chicago, IL 60637, USA}
\affil{Kavli Institute for Cosmological Physics, University of Chicago, Chicago, IL 60637, USA}

\author[0000-0001-5143-1255]{A. Chaturvedi}\email{aa07223@surrey.ac.uk}
\affiliation{Department of Physics, University of Surrey, Guildford GU2 7XH, UK}

\author[0000-0002-1763-4128]{D.~Crnojevi\'c}\email{dcrnojevic@ut.edu}
\affiliation{Department of Physics \& Astronomy, University of Tampa, 401 West Kennedy Boulevard, Tampa, FL 33606, USA}

\author[0000-0003-0105-9576]{G. E. Medina}\email{gustavo.medina@utoronto.ca}
\affiliation{David A. Dunlap Department of Astronomy \& Astrophysics, University of Toronto, 50 St George Street, Toronto ON M5S 3H4, Canada}
\affiliation{Dunlap Institute for Astronomy \& Astrophysics, University of Toronto, 50 St George Street, Toronto, ON M5S 3H4, Canada}

\author[0000-0001-5805-5766]{A. H. Riley}\email{alexander.riley2@durham.ac.uk}
\affiliation{Institute for Computational Cosmology, Department of Physics, Durham University, South Road, Durham DH1 3LE, UK}

\author[0000-0003-2497-091X]{N. Shipp}\email{nshipp@uw.edu}
\affiliation{Department of Astronomy, University of Washington, Seattle, WA 98195, USA}

\author[0000-0003-4341-6172]{A.~K.~Vivas}\email{kathy.vivas@noirlab.edu}
\affiliation{Cerro Tololo Inter-American Observatory/NSF NOIRLab, Casilla 603, La Serena, Chile}

\author{M. Wertheim}\email{maia.wertheim@mail.utoronto.ca}
\affiliation{David A. Dunlap Department of Astronomy \& Astrophysics, University of Toronto, 50 St George Street, Toronto ON M5S 3H4, Canada}
\affiliation{Dunlap Institute for Astronomy \& Astrophysics, University of Toronto, 50 St George Street, Toronto, ON M5S 3H4, Canada}

\author[0000-0003-1680-1884]{Y.~Choi}\email{yumi.choi@noirlab.edu}
\affiliation{NSF NOIRLab, 950 N. Cherry Ave., Tucson, AZ 85719, USA}

\author[0000-0002-9144-7726]{C.~E.~Mart\'inez-V\'azquez}\email{clara.martinez@noirlab.edu}
\affiliation{NSF NOIRLab, 670 N. A'ohoku Place, Hilo, Hawai'i, 96720, USA}

\author[0000-0001-9649-4815]{B.~Mutlu-Pakdil}\email{Burcin.Mutlu-Pakdil@dartmouth.edu}
\affiliation{Department of Physics and Astronomy, Dartmouth College, Hanover, NH 03755, USA}

\author[0000-0001-9438-5228]{M.~Navabi}\email{m.navabi@surrey.ac.uk}
\affiliation{Department of Physics, University of Surrey, Guildford GU2 7XH, UK}

\author[0000-0002-1594-1466]{J.~D.~Sakowska}\email{jsakowska@iaa.es}
\affiliation{Department of Physics, University of Surrey, Guildford GU2 7XH, UK}
\affiliation{Instituto de Astrofísica de Andalucía, CSIC, Glorieta de la Astronom\'\i a,  E-18080 Granada, Spain}

\author[0000-0003-1479-3059]{G.~S.~Stringfellow}\email{Guy.Stringfellow@colorado.edu}
\affiliation{Center for Astrophysics and Space Astronomy, University of Colorado Boulder, Boulder, CO 80309, USA}

\author[0000-0001-6455-9135]{A.~Zenteno}\email{alfredo.zenteno@noirlab.edu}
\affiliation{Cerro Tololo Inter-American Observatory/NSF NOIRLab, Casilla 603, La Serena, Chile}

\collaboration{25}{(DELVE Collaboration)}



\begin{abstract}

Extra-tidal features around globular clusters (GCs) are tracers of their disruption, stellar stream formation, and their host's gravitational potential.
However, these features remain challenging to detect due to their low surface brightness.
We conduct a systematic search for such features around 19 GCs in the DECam Local Volume Exploration (DELVE) survey Data Release 2, discovering a new extra-tidal envelope around NGC 5897 and find tentative evidence for an extended envelope surrounding NGC 7492.
Through a combination of dynamical modeling and analyzing synthetic stellar populations, we demonstrate these envelopes may have formed through tidal disruption.
We use these models to explore the detectability of these features in the upcoming Legacy Survey of Space and Time (LSST), finding that while LSST's deeper photometry will enhance detection significance, additional methods for foreground removal like proper motions or metallicities may be important for robust stream detection. 
Our results both add to the sample of globular clusters with extra-tidal features and provide insights on interpreting similar features in current and upcoming data.

\end{abstract}


\section{Introduction} 
\label{sec:intro}

Understanding the Milky Way's gravitational potential is key to answering several fundamental questions about the Local Universe  \citep{bg16}.
In particular, it largely reflects the distribution of dark matter at a given location and point in time.
Consequently, it dictates the motion of objects within the virial radius of our Galaxy, including stars \citep[e.g.,][]{xrz+08, dbe+12, ksl+12, ksl+14}, star clusters \citep[e.g.,][]{eh16, bw17, swf+18, wvr+19}, stellar streams \citep[e.g.,][]{krh10,nwy+10, gbe14,im+24}, and satellite galaxies \citep[e.g.,][]{we99, wea10,ccd+19}, along with perturbations from the Large Magellanic Cloud (LMC) and other prominent substructures \citep[e.g.,][]{dbl+19, slp+19, ebp+20, pel+22, kel+23}. 
This makes the Milky Way's gravitational potential important in understanding the assembly history of our Galaxy \citep[e.g.,][]{kis+22}, determining whether the LMC is on first-passage \citep[e.g., ][]{kvb+13,v+24}, and the infall and orbits of the Milky Way's satellite galaxies in the outer halo \citep[e.g.,][]{rpb+12, swt+24, ocs+24}; in addition to inherent questions about the shape of the dark matter halo of our Galaxy \citep[e.g.,][]{lm+10, hch+23, lpe+23}.

Gravitational potentials are notoriously difficult to measure since the mass budget of galaxies is dominated by dark matter \citep[e.g.,][]{rft+80, Helmi08}.
Ideally, 6D kinematics of luminous tracers (e.g., field stars, stellar streams, star clusters, dwarf galaxies) are required to model our Galaxy's underlying mass distribution \citep[e.g.,][]{jzs+99, pk13, iwt+13, vasiliev19, kel+23, bp24}. 
The past decade has seen significant advances in obtaining 6D kinematics of individual stars due to the success of the Gaia astrometry mission \citep{gaia+16} and targeted spectroscopic campaigns \citep[e.g.,][]{lkz+19, ljp+22}.
Notably, in the most recent data release (\textit{Gaia DR3}; \citealt{gaia+23}), the Gaia Collaboration published parallaxes and proper motions for $\approx 1.4 \times 10^9$ stars, and radial velocities for $\approx 10^7$ nearby luminous stars, completing the set of 6D kinematics for the latter subset.
While this is a welcome advance, it remains challenging to obtain constraints on the potential with these stars, since they are relatively local and therefore do not provide enough information about the potential at larger Galactocentric distances.
Using groups of stars (e.g., star clusters) is more promising as one can obtain distances from fitting isochrones to color-magnitude diagrams (CMDs), systemic radial velocities from bright members, and then use evidence of ongoing tidal disruption to probe the Galactic potential.
Notably, more than $150$ globular clusters (GCs) and dozens of dwarf galaxies \citep{h+96, vb+21, p24} are known around the Milky Way, extending well into its outer halo.

In addition to intact structures, ``streams" of stars from tidally disrupted (or currently disrupting) globular clusters and dwarf galaxies have been particularly powerful luminous tracers of the gravitational potential of the Milky Way and the LMC (see \citealt{johnston16, bp24} for recent reviews, and references therein).
The large spatial extents of streams \citep[e.g.,][]{mateu+23}, and the fact that they roughly correspond to past orbits, make them particularly powerful probes of wide extents of the Milky Way's gravitational potential \citep[e.g.,][]{kel+23}. 
Additionally, the dynamically fragile nature of GC streams makes them promising tools to search for interactions with low mass dark subhalos predicted by the $\Lambda$CDM model \citep[e.g.,][]{ebb+16}.
These results have collectively motivated future searches for streams as a priority for the upcoming Legacy Survey of Space and Time (LSST) with the Vera Rubin Observatory \citep{ikt+19}, and highlight the importance of finding more of these structures in the Milky Way.

A promising avenue to increase the population of objects that are currently disrupting and producing streams is to search for extra-tidal features around globular clusters.
The Milky Way's GCs are old ($> 10$\,Gyr), self-gravitating, densely populated objects that are orbiting our Galaxy.
As noted, the Milky Way has $>$150 known GCs, of which multiple have prominent extra-tidal features (e.g., ``envelopes", or streams) due to their interaction with the Milky Way's gravitational potential (e.g., \citealt{ogr+01, spr+03, jg+10, bsd+11, smv+11, cs+16, sdb+18, mcs+18b, khw+19, c+19, g+19, s+20, csp+20, pmc+21, bnc+21, imm+21, pm+21, kns+22,zmd+22} and \citealt{mateu+23} for a recent compilation and references therein).
However, extra-tidal features around GCs are still difficult to detect, due to their generally low surface brightnesses ($\sim$31\,mag\,arcsec$^2$; e.g., \citealt{smm+18}).
Despite this challenge, techniques such as matched-filter searches using color-magnitude information have uncovered tidal features around many GCs \citep{bei+06, bsd+11, nbk+17}, including the famous $\sim 30 \deg$ tidal tails around Palomar~5 \citep[e.g.,][]{rog+02, bpp+20}.
While extended features like these are the most constraining in terms of the Milky Way potential, the discovery of any tidal deformations (e.g., small envelopes) may indicate the presence of even fainter stellar streams potentially detectable by upcoming surveys.

This project adds to the list of such studies by searching the regions around GCs located in Data Release 2 of the DECam Local Volume Exploration survey \citep[DELVE DR2;][]{dfa+22} footprint for any associated extra-tidal features.
Our results largely replicate those of previous studies \citep[e.g.,][]{zmd+22}, but we identify one GC (NGC 5897) with evidence for an extra-tidal envelope that has not previously been reported, in addition to another GC (NGC 7492) with tentative evidence.
We then seek to understand whether this detection may reflect fainter tidal features by simulating the disruption of NGC 5897 using the \texttt{gala}\footnote{https://gala.adrian.pw/en/latest/} python Galactic dynamics package \citep{gala}, along with the GCs Pal 5 (prominent tidal tails) and NGC 5634 (no extra-tidal features) for comparison.
We then generate simulated deeper data in these GCs to mimic future LSST observations, including stellar population models (SPISEA; \citealt{hll+20}) and Milky Way foreground contamination (LSST SIM DR2\footnote{https://datalab.noirlab.edu/lsst\_sim/}; \citealt{ggh+05, dpm+22}), in order to forecast whether our tentative extra-tidal envelopes will be clearly detectable and indicate underlying stellar streams.

This paper is organized as follows. 
In \secref{obs}, we describe the DELVE survey and the way we searched for extra-tidal features. 
In \secref{results}, we present our finding for extra-tidal features around GCs in the DELVE footprint.
In \secref{modeling}, we detail our tidal disruption simulation method.
Finally, we compare the results of our simulations to observations and look ahead to LSST in \secref{modeling_analysis} before concluding in \secref{conclusion}.

\section{Data \& Observational Analysis} \label{sec:obs}

We conducted a search for extra-tidal features in the surrounding area of MW GCs (out to 2$^\circ$ from their centers in projected distance) in the DELVE DR2 footprint with reasonably complete $g, r$ photometry coverage. 
Here we describe the DELVE DR2 dataset from which we used photometry to search for extra-tidal features (Section~\ref{sec:delve}), the Gaia DR3 proper motions and parallaxes which we used to exclude non-members (Section~\ref{sec:gaia}), and the procedure we used to search for overdensities beyond the tidal radius of GCs (Section~\ref{sec:obs_analysis}). 

\begin{figure*}[th!]
    \centering
    \includegraphics[width=\linewidth]{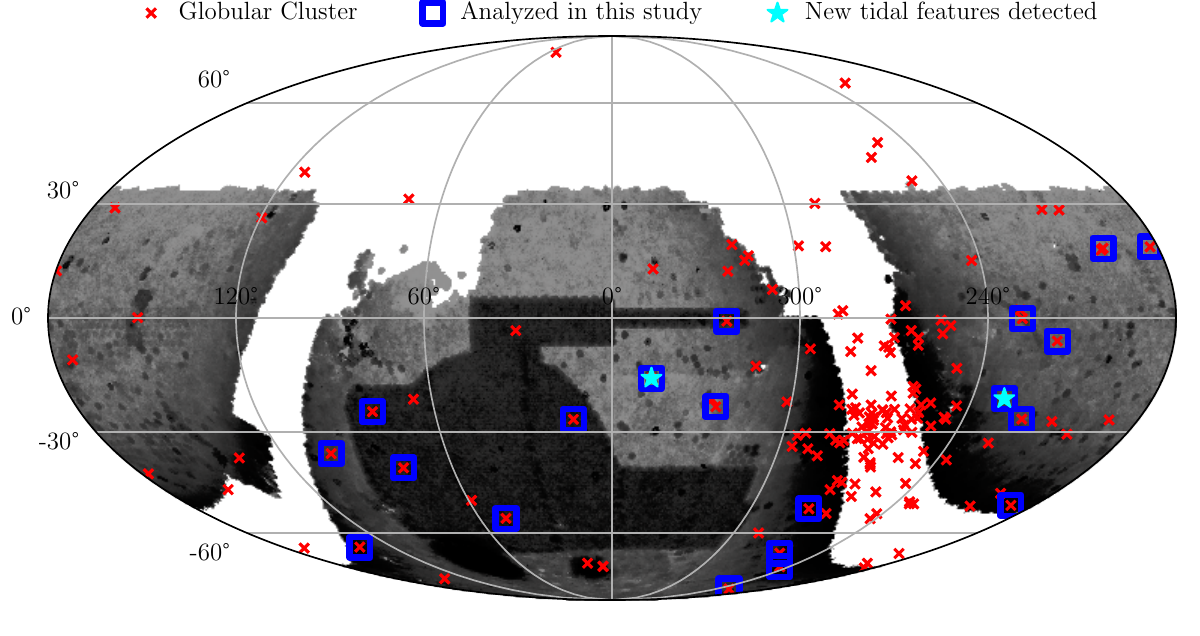}
\caption{The density of sky coverage, plotted in equatorial coordinates (RA and Dec), with the DECam camera and the location of 170 globular clusters (GCs) from \citet{vb+21} and the overlap with the DELVE footprint. The Dark Energy Survey (DES) covers the irregular polygon of higher star density in the lower center of the figure. 
The rest of the coverage is from the DELVE survey. 
GCs are denoted with a red cross, and those which had enough high quality data that we could search for extra-tidal envelopes have a blue square around the cross. 
NGC 5897 (the GC with a newly detected envelope) and NGC 7492 (which has conflicting reports in the literature and for which we detect an envelope), are shown as cyan stars with coordinates $(\alpha = 229.35\degree, \delta = -21.01\degree)$ and $(\alpha = 347.11\degree, \delta = -15.61\degree)$, respectively.}
\label{fig:delve_footprint}
\end{figure*}

\subsection{DELVE DR2}
\label{sec:delve}

DELVE DR2\footnote{https://datalab.noirlab.edu/data/delve} provides coverage of $\sim21,000$ sq. degrees of the high-latitude ($|b| > 10^\circ$) sky in the broadband DECam $g,r,i,z$ filters, providing a high-quality dataset for photometric studies of Milky Way substructures \citep{dfa+22}.
DELVE aims to combine archival Dark Energy Camera (DECam; \citealt{fdh+15}) observations with more than 150 nights of new data for contiguous coverage of the high-latitude southern sky. 
DELVE DR2 includes $>160,000$ exposures from a variety of NOIRLab community programs, with the largest number of exposures from DELVE \citep{dcn+21}, DES \citep{DES+21}, DECaLS \citep{dsl+19}, and the DECam eROSITA Survey \citep[DeROSITAS;][]{zkk+25}.
Exposures are uniformly processed by the DES Data Management (DESDM; \citealt{mgm+18}) pipeline, and resulting source catalogs are merged across exposures following \citet{dbm+20}.
The resulting median 5\,$\sigma$ source depth for each band is $g = 24.3, r = 23.9, i = 23.5,$ and $z = 22.8$ (see Figure 10 in \citealt{dfa+22}).

The data in DELVE DR2 is an amenable dataset to search for previously undetected extra-tidal features in regions immediately surrounding GCs, motivating our particular analysis.
First, DELVE DR2 includes all suitable public archival data from DECam at the point of its processing ($> 270$ community programs; \citealt{dfa+22}), providing a unique catalog for photometric studies. 
Second, it is easier to account for the effects of systematic variations (e.g., large-scale inhomogeneities in exposure depth, coverage) in searches for overdensities in the near-vicinity of GCs, which is important given that DELVE DR2 does not uniformly cover the southern hemisphere.
Furthermore, the dataset has been used very successfully for the discovery and characterization of other faint, metal-poor systems in the Milky Way (e.g., \citealt{mcp+20, cpd+21, psd+22, csl+23, cmd+23, ccg+25, tcd+25}).
Motivated by these factors, we queried DELVE DR2 to create catalogs with all stars within a $4 \degree \times 4 \degree$ region around each GC in \citet{h+96} at high Galactic latitude ($|b| > 10^\circ$) and low declination $\delta < 30^\circ$. 
We performed a matched filter analysis based on isochrones with broadband $g,r$ photometry.
In each $4 \degree \times 4 \degree$ area, which corresponded to physical projected widths of $\sim$0.5\,kpc to $\sim$2.5\,kpc depending on the GC distance, we reached depths of $g\gtrsim 23$.
We also required a relatively constant depth, along with a clearly sampled color-magnitude sequence.
This left us with 19 GCs out of an initial 57 for our analysis, out to distances of $\sim35\,$kpc (see Table~\ref{tab:table_results}).
The coverage of DELVE DR2 and the location of known GCs are shown in Figure~\ref{fig:delve_footprint}.

The initial data cleaning on the resulting catalogs is as follows. 
First, we de-reddened the data using the \texttt{extinction\_\{g,r,i\}} columns in DELVE DR2, following the reddening corrections derived in \citet{dfa+22}.
Next, we performed an initial quality check by excluding all sources with \texttt{flags\_\{g,r\}} $\ge$ 4 and \texttt{magerr\_psf\_\{g,r\}} $\ge$ 0.5, to exclude sources that are saturated or have very large photometric uncertainties.
Then, we performed a star/galaxy separation by requiring $|$\texttt{spread\_model\_g}$|$ $< 0.002\,+\,3/5\,\times\,$\texttt{spreaderr\_model\_g}, where \texttt{spread\_model\_g} and \texttt{spreaderr\_model\_g} come from a likelihood-based star-galaxy classifier defined in \citet{dam+12}.
The DELVE DR2 dataset also provides maps of the effective depth from the effective-exposure-time scale factor ($t_{\text{eff}}$; \citealt{nbg+16}) as a function of location on the sky in DECam $g,i$.
These maps were used to visually flag over-densities that were correlated with these $t_{\text{eff}}$ maps as spurious detections. 

We then estimated the median depth at which the source catalog was $\sim50\,\%$ complete within a $2^\circ$ radius around each GC.
This was performed using the DELVE DR2 \texttt{maglim} survey property maps (see Appendix E of \citet{Sevilla-Noarbe:2021} for more details) in these regions to derive the median 5\,$\sigma$ depth, and estimating the depth at 50\,\% source completeness in the $g$ band using $g_{50} = 1.14\,\times\,g_{5\sigma} - 4.51$ and in the $r$ band $r_{50} = 1.08\,\times\,r_{5\sigma} - 3.23$.
These empirical equations were derived by performing a crossmatch between DELVE DR2 and the Hyper Suprime-Cam Subaru Strategic Program Public Data Release 3 \citep[HSC-SSP PDR3; ][]{Aihara:2022} for sources classified as stars. 
Assuming the HSC-SSP is complete at DELVE magnitudes, we compute the magnitude at which DELVE completeness, compared to HSC-SSP, equals 50\,\% at a resolution of \texttt{NSIDE=128} ($\sim6.4\times10^{-5}$ steradians). 
The above equations are then derived by fitting a line in each band to obtain 50\,\% completeness as a function of the \texttt{maglim} survey property.
We excluded sources fainter than these depths to ensure reasonably complete and homogeneous catalogs in our search for extra-tidal features.

\subsection{Gaia DR3}
\label{sec:gaia}

We supplemented our photometric dataset with astrometric data from \textit{Gaia} DR3 \citep{gaia+16,lkh+21,gaia+23} at the bright-end of our catalog ($g \lesssim 20.5$) to exclude stars with proper motions or parallaxes inconsistent with GC membership.
Specifically, {\it Gaia} DR3 provides proper motions with uncertainties of $\sim0.5$\,mas\,yr$^{-1}$ at $G = 20$ and parallaxes with uncertainties of $\sim0.07$\,mas when $G=17$\footnote{https://www.cosmos.esa.int/web/gaia/dr3}.
To incorporate this information in our search for extra-tidal features, we first cross-matched our DELVE DR2 query around each GC with \textit{Gaia} DR3 and compiled any available astrometric and proper motion information for each source.
As described further in \secref{obs_analysis}, we then excluded stars from our matched-filter analysis if they had proper motions more than 3$\sigma$ discrepant from the systemic proper motion of each GC. 
In this exclusion of non-members, 1$\sigma$ is defined as each star's proper motion uncertainty added in quadrature to a conservative 4\,km\,s$^{-1}$ internal dispersion (see Appendix plots in \citealt{vb+21}).
In addition, for each star with a significantly resolved parallax in \textit{Gaia} (\texttt{parallax\_over\_error > 5}), we excluded the source if the median geometric distance estimate of the star in the \textit{Gaia} distance catalog of \citet{brf+21} was 3$\sigma$ discrepant from the distance of the GC in \citet{h+96} (\ie $|\texttt{r\_med\_geo} - \textrm{dist}_\textrm{GC}| < 3 \times (\texttt{r\_hi\_geo} - \texttt{r\_lo\_geo})/2$).
These two selections ensured that the bright end of our catalog in the matched-filter analysis was not heavily contaminated by Milky Way foreground stars.
Naturally, we did not apply any of these selections for stars with no entries in \textit{Gaia} DR3 (e.g., $g \gtrsim 20.5$).

\subsection{Search for GC extra-tidal over-densities}
\label{sec:obs_analysis}

\begin{figure*}[th!]
    \centering
    \includegraphics[width=\linewidth]{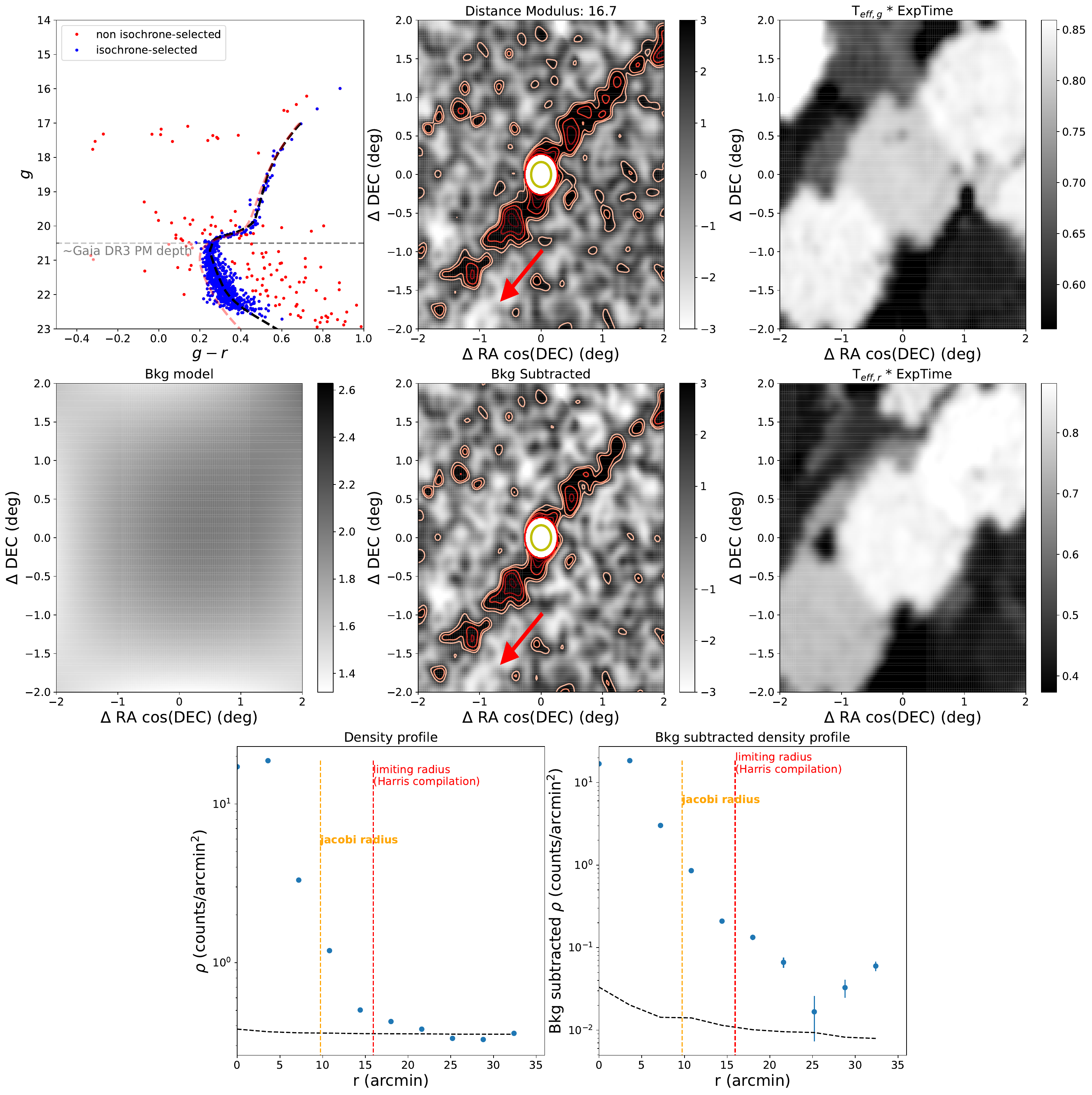}
\caption{Top left: A color-magnitude diagram of the 1000 closest stars to the center of Pal 5, after excluding those with bad photometry (Source Extractor parameter \texttt{FLAGS $>$ 3}; see Section~\ref{sec:delve}). 
A cubic spline interpolation along the CMD of the cluster is shown as a dashed black line, and sources with photometry consistent with being within 0.05\,mag of this fitted function are in blue.
The rough depth of usable Gaia DR3 proper motions ($g\sim20.5$) is indicated with a dashed grey line.
The MIST isochrone with the closest age and metallicity to the values in the \citet{h+96} catalog is shown as a light red line.
Top middle: A density map of all sources within a 4$^\circ$ box around Pal 5, scaled by the standard deviation of the source density. 
Contours correspond to 2, 3, 5, 7, 10, 50, 100 sigma overdensities. 
The central King limiting radius $r_t$ of the GC is masked and circled in red, and the orange circle corresponds to the Jacobi radius calculated in this work (Section~\ref{sec:obs_analysis}).
The red arrow indicates the direction of the systemic proper motion of Pal 5.
Top right: A map of the effective exposure time ($t_{\text{eff}} \times T_{exp}$) in DELVE DR2 for the $g$ band photometry over the same region as the middle panels.
Middle left: A background model for the top middle panel, derived by fitting a second-order two-dimensional polynomial between $>$4\,$r_t$ and the entire 4$^\circ \times 4^\circ$ field-of-view.
Middle: Background-subtracted source density plot.
Middle right: A map of the effective exposure time ($t_{\text{eff}} \times T_{exp}$) in DELVE DR2 for the $r$ band photometry over the same region as the middle panels.
Bottom left: Angular-averaged source density plot of the top middle panel, with the limiting radius derived from \citet{h+96} marked in red, and the Jacobi radius calculated in Section~\ref{sec:obs_analysis} in yellow. 
The error bars on the data points are the standard error in the spatial bins averaged (1.8$'$) over the annulus, in steps of 3.6$'$.
The dashed line corresponds to the background value plus the standard error.
Bottom right: Same as bottom left, but for the background-subtracted density plot.}
\label{fig:pal5example}
\end{figure*}

We performed a search for photometric overdensities around each GC as follows.
We retrieved isochrones from the MESA Isochrones and Stellar Tracks (MIST) library   \citep{cdc+16, d+16, pbd+11, pca+13, pms+15, psb+18} closest to the age and metallicity of each GC based on parameters in the \citet{h+96} catalog.
We then placed the isochrone at the distance modulus of the GC, based on values in \citet{bv+21}, and retained all stars with $g,r$ photometry that was 2$\sigma$  consistent (as per the photometric uncertainties) with a 0.05\,mag color tolerance around the isochrone.
Following this step, we computed the systemic proper motion of the GC by fitting a 2D Gaussian to the proper motion distribution of stars within its King limiting radius that passed the isochrone selection. 
We note that these values were consistent with the systemic proper motions reported in the literature (e.g., \citealt{vb+21}).
We then excluded stars that had proper motion entries in \textit{Gaia} DR3 that were inconsistent at the 3$\sigma$ level from the derived systemic proper motion, and also excluded those with resolved parallaxes following Section~\ref{sec:gaia}.
Density maps of the resulting source catalog were then generated using a pixel scale of 1.8\arcmin\,\,and smoothed with a Gaussian kernel of $\sigma =$ 3.6\arcmin, at distance moduli spanning $\pm$1\,mag around the value for the GC in 0.2\,mag increments.
This $\sigma$ value was chosen because it is $0.2-0.4\times$ the King limiting radius (also referred to in the literature as the King ``tidal" radius $r_t$) of the GCs we searched, providing a reasonable visual density map that could reveal small-scale features in the near vicinity.
In the density maps, stars without proper motions are downweighted by the fraction of stars that were excluded by the proper motion selection, to account for their higher likelihood to be foreground stars.
These King limiting radii were calculated from the concentration and core radii in the updated \citet{h+96} compilation.
Note that this limiting radius is a structural parameter of the GC, and is different from the Jacobi radius ($r_J$) which defines the distance out to which stars are gravitationally bound to the host GC.
To account for spatial incompleteness of the survey, the pixelized density maps were divided by the fraction of each pixel that was observed in DELVE-DR2, which is the same method used in \citet{sdb+18} and \citet{psd+22}.
Pixels with an observed fraction less than 0.5 were excluded from this analysis.

To interpret the resulting density plots and have an initial assessment of the robustness of extra-tidal overdensities around these GCs, we generated eight-panel plots of which Figure~\ref{fig:pal5example} is an example for Pal 5. 
A detailed description of the panels is given in the caption for Figure~\ref{fig:pal5example}, but we describe several relevant panels here. 
The top center panel is the raw density map discussed by the previous paragraph, where the colorbar ranges from $\pm3\sigma$ where $1\sigma$ corresponds to a sigma-clipped standard deviation of the fluctuations outside of $4\,r_t$.
Note that the region of the cluster within the King limiting radius ($r_t$) is masked to increase the dynamic range of the density plot. 
The center left panel is a background estimate derived from fitting a second-order two-dimensional polynomial to the density map outside of $4\,r_t$ in the center top panel, and the center panel is the background-subtracted density map.

The Jacobi radius $r_J$ is overplotted as a yellow circle, and we consider overdensities outside this region to be ``extra-tidal."
We get $r_J$ from equation 7 of \citet{k62}
\begin{equation}
    r_J = \bigg(\frac{Gm_{\textrm{GC}}}{\omega^2 - \frac{d^2\Phi_G(R)}{dR^2} \Bigr|_{R_{\textrm{GC}}}}\bigg)^{1/3} \label{eq:tidal_radius}
\end{equation}
In this construction, $\omega$ is the angular velocity, $m_{\textrm{GC}}$ is the mass of the GC, $R_{\textrm{GC}}$ is the distance of the GC from the Galactic center, and $\Phi_G$ is the Milky Way potential, for which we use \texttt{MilkyWayPotential2022} from Gala, which is an excellent match to observations (see \secref{simdebris} for more detail on this potential).
Formally, this definition for $r_J$ assumes that the gravitational potential is spherical, which is not strictly the case but ought to be a reasonable approximation.
This formalism differs slightly from another commonly used definition $r_J = R_{\textrm{GC}}(m_{\textrm{GC}}/2M)^{1/3}$ \citep[][their equation 3]{k62}, which assumes a flat rotation curve for the Galaxy.
In Table~\ref{tab:table_results}, we report both our derived $r_J$ and the $r_J$ reported by \citet{bg+18} which uses the \citet{k62} formula.
We note that the $r_J$ at pericenter is also used to motivate whether objects ought to be stripping (e.g., \citealt{pel+22}), so we include it in Table~\ref{tab:table_results} for reference.

To ensure density fluctuations are not caused by systematics in coverage, we plot maps of the $t_{\text{eff}}$ parameter multiplied by the exposure time for $g,r$ in the top and center right panels (see Section~\ref{sec:delve} for details on $t_{\text{eff}}$) as a check of whether these tightly track any overdensities. 
We note that in Figure~\ref{fig:pal5example}, the $t_{\text{eff}}$ maps track the Pal 5 tails since the NOIRLab archive contains deeper data that was obtained around them following their discovery. 
Finally, we overplot radial density profiles of the GC with and without background subtraction in the bottom left and right panels, respectively, to assess the significance of extra-tidal overdensities.
For NGC 5897 and NGC 7492, the two GCs for which we claim extra-tidal detections that are new or have conflicts in the literature, we further statistically assess the significance through Monte Carlo resampling the noise in the images (see Sections~\ref{sec:ngc5897_obs} and~\ref{sec:ngc7492}).

We note that after our initial tentative new detections, we further refined the model isochrones in all GCs to ensure a closer match to their stellar population.
Specifically, we interpolated along the color-magnitude distribution of the 1000 stars closest to the center of each GCs to derive an empirical isochrone, and then repeated the analysis. 
We opted for this approach as opposed to adjusting the parameters of the isochrones, as minor parameter adjustments did not fully fix misalignments.
The isochrones were derived using \texttt{scipy.interpolate.LSQUnivariateSpline} with custom knot spacing, and are overplotted in the top left panels of Figures~\ref{fig:pal5example},~\ref{fig:ngc5897}, and~\ref{fig:ngc7492}.
The collection of eight panel plots for our sample of GCs in Table~\ref{tab:table_results}, including those with no extra-tidal features, are hosted in an online Zenodo repository at 10.5281/zenodo.15045876.

\section{Observational Results} \label{sec:results}

Our search for extra-tidal features around 19 GCs in DELVE DR2 largely replicates results from existing compilations \citep[\eg][]{pc20, zmd+22}, but reveal evidence for potentially new extra-tidal features around one GC (NGC 5897; see Figures~\ref{fig:ngc5897}) and tentatively for another with contradictory evidence in the literature (NGC 7492; see Figure~\ref{fig:ngc7492}).
The additional sample of GCs investigated for extra-tidal features constitute $\sim 20\%$ of the known high-latitude ($|b|>10\deg$) GCs and was set by the quality (e.g., homogeneity, coverage) of the matched-filter maps described in Section~\ref{sec:obs_analysis}.
We list these GCs and our results for each in Table~\ref{tab:table_results}.

When comparing our results to the literature, we stress that different methods have been used to study these systems.
These methods often search for entirely difference signatures of tidal disruption.
For example, STREAMFINDER \citep{mi18} looks for elongated structures spanning tens of degrees, while we search within $\sim 2 \degree$ of the progenitor.
Therefore, our non-detections of extended extra-tidal tails should not refute prior work.
We point to NGC 1904 as one example where our search in the near vicinity of the GC shows possible but inconclusive signs of tidal disruption, but previous wider area searches find more definitive signs of a stream \citep{sdb+18, ale+25}.
We now discuss the new detection of extra-tidal features in NGC 5897 (Section~\ref{sec:ngc5897_obs}), along with our analysis for NGC 7492 (Section~\ref{sec:ngc7492}).

\begin{table*}[h]
\centering
\begin{adjustbox}{angle=90, center}
\begin{minipage}{\textheight}
\begin{tabular}{c c c c c c c c c}
    Name & R.A. (\degree) & Dec (\degree) & $d$ (kpc) & $r_J$ (pc) & $r_{J,\textrm{peri}}$ (pc) & $r_{J,\textrm{King}}$ (pc) & This study &  Previous detections \\
    \hline
    NGC 288 & 13.2 & $-26.59$ & 9.0 & 96.01 & 71.87 & 76.43 & E & S18(T), K19(T) \\
    NGC 1261 & 48.06 & $-55.22$ & 16.4 & 133.32 & 71.61 & 146.38 & N?$^a$ & S18(T), I21(T), A25(T) \\
    NGC 1851 & 78.53 & $-40.05$ & 12.0 & 151.94 & 71.17 & 166.46 & E & S18(T), I21(T) \\ 
    NGC 1904 / M79 & 81.04 & $-24.52$ & 12.9 & 140.89 & 62.39 & 153.79 & N?$^b$ & contradictory (S18(T), Z22(N), A25(T)) \\
    NGC 2298 & 102.25 & $-36.01$ & 10.8 & 49.99 & 30.26 & 81.27 & E & contradictory (I21(T), Z22(N)$^c$) \\
    NGC 2808 & 138.01 & $-64.86$ & 9.6 & 157.12 & 86.54 & 176.87 & N  & I21(T) \\
    NGC 4147 & 182.53 & 18.54 & 19.0 & 86.78 & 48.85 & 95.97 & E & JG1(T) \\
    C1235-509 / Rup 106 & 189.67 & $-51.15$ & 21.2 & 127.81 & 92.42 & 96.01 & N & None \\
    NGC 5024 / M53 & 198.23 & 18.17 & 18.4 & 183.80 & 160.65 & 199.02 & N & B21 (association with streams) \\
    NGC 5634 & 217.41 & $-5.98$ & 25.3 & 175.71 & 120.66 & 158.46 & N & K22 find 10 stars outside $r_J$ \\
    NGC 5694 & 219.9 & $-26.54$ & 33.9 & 263.98 & 163.69 & 203.68 & N & M18(E) \\
    IC 4499 & 225.08 & $-82.21$ & 18.4 & 119.29 & 100.93 & 112.86 & N & None \\
    Pal 5 & 229.02 & $-0.11$ & 22.6 & 61.71 & 60.92 & 60.89 & T & R02, B20, and others \\
    \textbf{NGC 5897} & \textbf{229.35} & \textbf{$-$21.01} & \textbf{12.7} & \textbf{86.02} & \textbf{69.24} & \textbf{62.99} & \textbf{E} & \textbf{None$^d$} \\
    NGC 6101 & 246.45 & $-72.2$ & 15.1 & 91.04 & 90.55 & 78.02 & E & I21(T) \\
    NGC 6362 & 262.98 & $-67.05$ & 8.0 & 56.63 & 48.91 & 44.76 & N & S20(T) \\
    NGC 6584 & 274.66 & $-52.22$ & 13.2 & 56.36 & 44.21 & 69.47 & N?$^a$ & None \\
    NGC 7089 / M2 & 323.37 & $-0.82$ & 11.5 & 141.78 & 66.85 & 149.59 & E? & I21(T) \\
    \textbf{NGC 7492} & \textbf{347.11} & \textbf{$-$15.61} & \textbf{26.6} & \textbf{93.46} & \textbf{62.43} & \textbf{96.11} & \textbf{E?} & \textbf{contradictory (N17(T), Z22(N), I24(T))}

\end{tabular}
\caption{Properties of our studied GCs are shown here, including coordinates, heliocentric distance \citep{bh18}, Jacobi radii calculated in this paper ($r_J$; see Section~\ref{sec:obs_analysis}), Jacobi radii at pericenter ($r_{J,peri}$), and those in \citealt{bg+18} ($r_{J,\text{King}})$. 
In the ``This study" column, N denotes no extra-tidal feature, E denotes an extra tidal envelope and T denotes extra-tidal tails.
If a previous study found a different category of extra-tidal features than we do, we note their classification in parentheses in the ``Previous detections" column. 
Note that our study is biased toward finding envelopes as opposed to tidal tails due to our search being in the relative vicinity of each GC.
The full eight-panel figures for each of these GCs are hosted on Zenodo \citep{delvegcs}.
Legend for previous detections: S18 \citep{sdb+18}, K19 \citep{khw+19}, I21 \citep{imm+21},  \citep{zmd+22}, A25 \citep{ale+25}, JG10 \citep{jg+10}, B20 \citep{bpp+20}, \citep{bnc+21}, K22 \citep{kns+22}, M18 \citep{mcs+18a}, R02 \citep{rog+02}, S20 \citep{s+20}, N17 \citep{nbk+17}, I24 \citep{im+24}.\\
$^a$ NGC 1261 and NGC 6584 appear to be filled out to approximately their Jacobi radii.\\
$^b$ NGC1904 appears to have an extended feature in our analysis before background subtraction (see corresponding gif in the Zenodo repository at \citealt{delvegcs}).\\
$^c$ While \citet{zmd+22} find no evidence for extra-tidal features in NGC 2298, enough evidence exists in the literature \citep[e.g.,][]{s+20, imm+21} that they classify it as "E" in their Table~3.\\
$^d$ While there is no refereed article claiming a detection around NGC 5897, we note that an older conference proceeding \citep{dg06} claims the detection of tidal tails in a perpendicular direction to our detected envelope.
}
\label{tab:table_results}
\end{minipage}
\end{adjustbox}
\end{table*}

\subsection{NGC 5897} \label{sec:ngc5897_obs}

\begin{figure*}[th!]
    \centering
    \includegraphics[width=\linewidth]{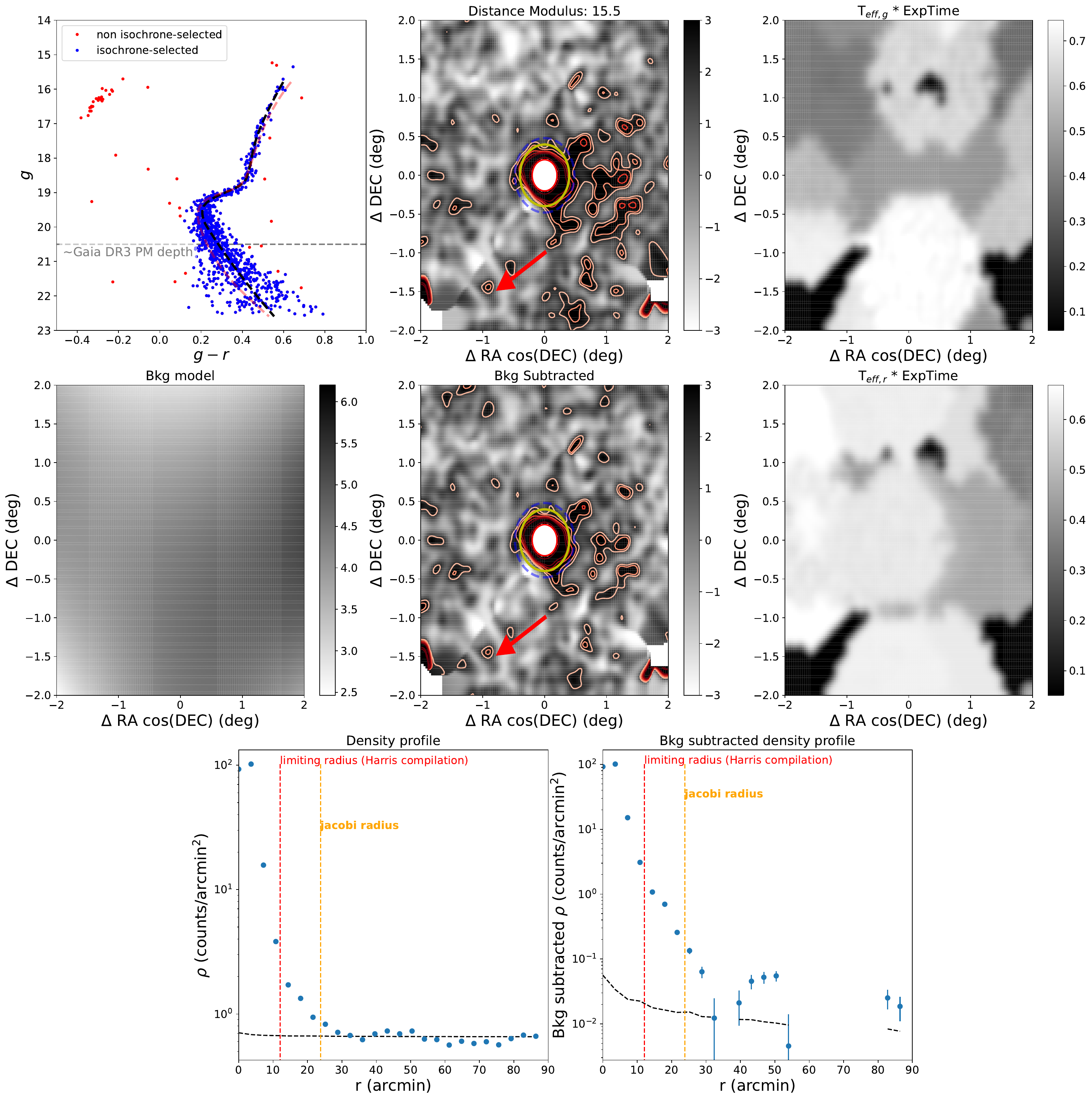}
\caption{Same as Figure~\ref{fig:pal5example}, but for NGC 5897. 
 A clear photometric overdensity is seen extending from the GC beyond $r_J$ (yellow circle). 
 A dashed blue circle is overplotted at $r=$ 28' to guide the eye, roughly at the radius out to which the photometric excess exists.
 Note that the excess does not preferentially align with the proper motion of NGC 5897 (red arrow); this is expected given the recent pericenter passage of NGC 5897 (see \ref{sec:ngc5897_obs} for more details).}
\label{fig:ngc5897}
\end{figure*}

\begin{figure}[th!]
    \centering
    \includegraphics[width=0.4\textwidth]{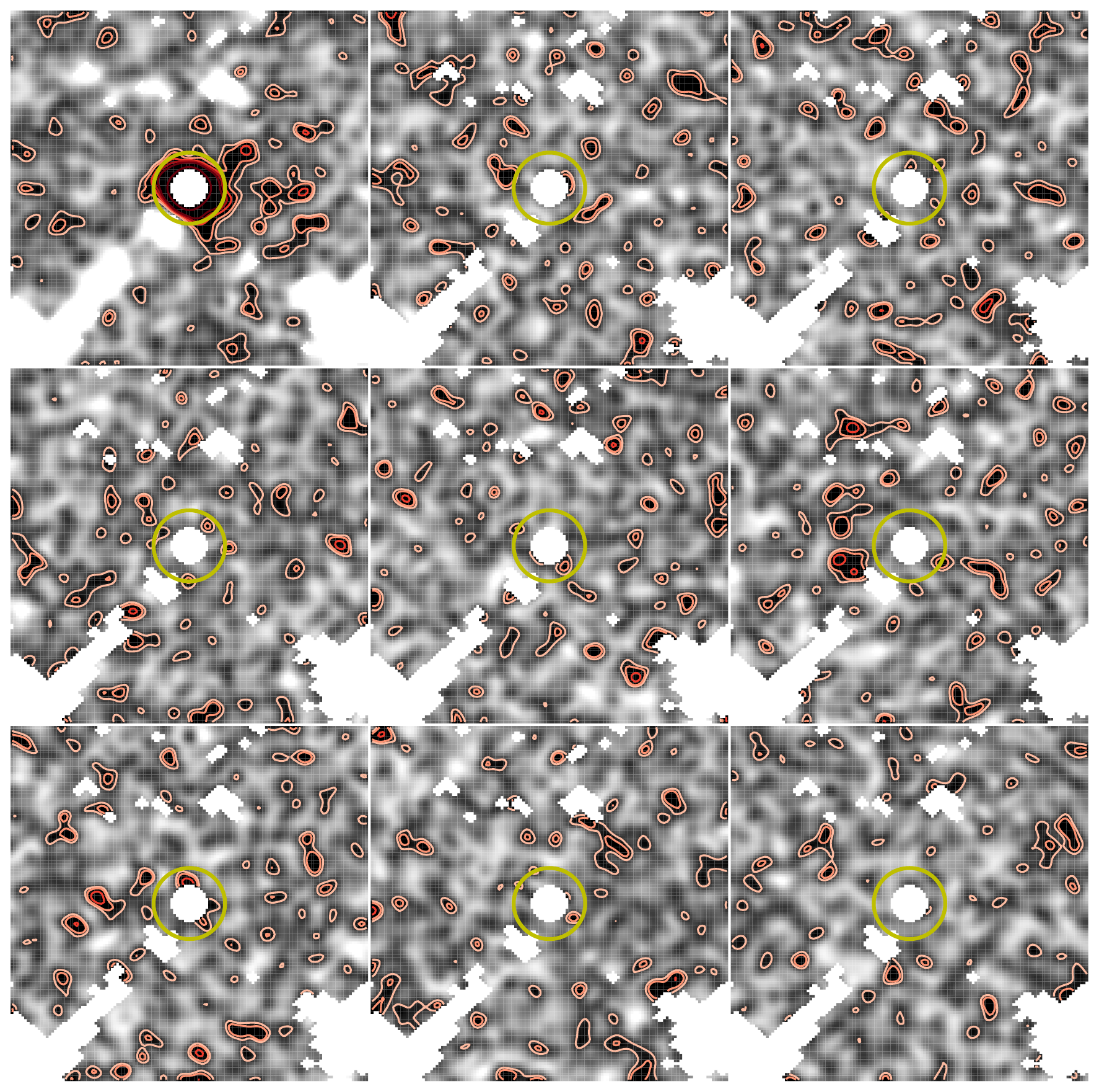}
    \includegraphics[width=0.4\textwidth]{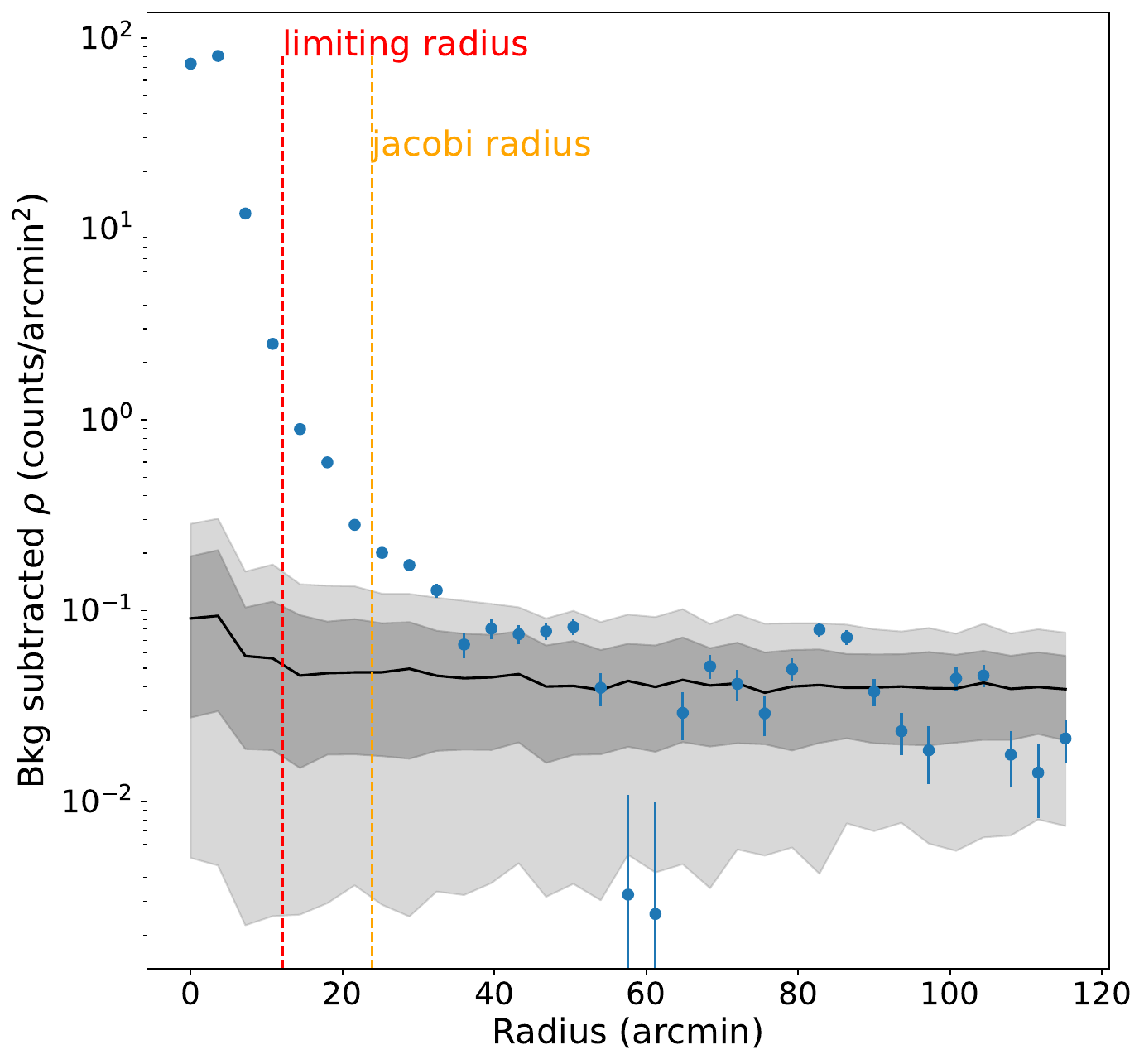}
\caption{Top: A 3x3 postage stamp plot where the top left panel is the observed density plot NGC 5897, and the other panels are example noise maps generated to assess the significance of the excess in NGC 5897, as described in Section~\ref{sec:ngc5897_obs}. 
Note that regions masked in white correspond to pixels with values equal to 0 in the raw data, corresponding to a lack of coverage (see Figure~\ref{fig:ngc5897}). 
The red and yellow circles correspond to $r_t$ and $r_J$, respectively, and the colorscale corresponds to what is shown in Figure~\ref{fig:ngc5897}. 
Bottom: An azimuthally averaged density profile of the background subtracted observed data is shown as blue points, with the mean, 1$\sigma$, and 2$\sigma$ of the distribution of density profiles from 1000 noise map samples shown in black/grey. 
The observed excess at the $r_J$ is at a 4.6\,$\sigma$ significance relative to distribution of density profiles from the noise maps.
}
\label{fig:ngc5897_mc}
\end{figure}

NGC 5897 ($\alpha = 229.35 \degree, \delta = -21.01 \degree$) is one of the two GCs around which we find new evidence for an extra-tidal envelope beyond $r_J$.
The matched-filter density plot that was used to diagnose the presence of the extra-tidal features is shown as the center panel of Figure~\ref{fig:ngc5897}.
We see a significant photometric excess beyond $r_J$ (yellow circle), making this a fairly confident detection of extra-tidal structure around NGC 5897. 
To statistically assess the significance of the extra-tidal overdensity, we perform a Monte Carlo analysis by generating noise maps by sampling the distribution of pixel values outside of 4\,$r_t$ (see top panel of Figure~\ref{fig:ngc5897_mc}). 
We then generate azimuthally averaged density profiles of 1000 of these noise maps and calculate the significance of the excess of the observed density profile at $r_J$ relative to this distribution of null density profiles, to derive a 4.6\,$\sigma$ significance for the detection (see bottom panel of Figure~\ref{fig:ngc5897_mc}).

The proper motion of NGC 5897 is shown as a red arrow in the center panel of Figure~\ref{fig:ngc5897}, and the photometric excess does not appear to preferentially lie along this direction.
In a recent analysis modeling other GCs, \citet{ate+25} found that a perpendicular feature could form from a recent pericenter passage, before the stars have the time to spread out along the sky, while extended tidal tails occur from previous passages.
We calculate the timing of NGC 5897's most recent pericenter passage by taking its current 6D phase space position from \citet{vb+21} and \citet{bh18} and backwards integrating its orbit within \texttt{MilkyWayPotential2022}.
We find it was most recently at pericenter $<50 \Myr$ ago, making this a plausible scenario.

In the literature, neither the recent \citet{zmd+22} nor \citet{s+20} compilations find evidence for extra-tidal features around NGC 5897, although an earlier conference proceeding does make a tentative claim of extra-tidal features \citep{dg06}.
Notably, the \citet{s+20} work is based on \textit{Gaia} DR2 magnitudes and proper motions and no previous published work exists on extra-tidal excesses around NGC 5897 based on deep, broadband photometry \citep{gd+04, dg06}.
Consequently, our discovery of extra-tidal features using wide-field photometry of this GC down to $g\sim22.7$ further highlights the discovery potential of wide-field, deep photometric surveys in systematic studies of previously well-known objects.

Notably, NGC 5897 is one of the most metal-poor GCs known ([Fe/H] $= -2.04$; \citealt{km+14}) and is located in the inner-halo of the Milky Way ($R_{GC} \approx 7.48$\,kpc; \citealt{bhs+19}).
Recent kinematic grouping analyses suggest that NGC 5897 may have formed in-situ (e.g., \citealt{bk+24, cg+24}), whereas other analyses aimed at classifying individual clusters suggest NGC 5897 is associated with the Gaia-Sausage-Enceladus merger \citep{mkh+19, ccd+22}.
Since NGC 5897 is constrained to the inner halo \citep[$R_{\textrm{peri}} = 2.86$\,kpc, $R_{\textrm{apo}} = 9.31$\,kpc;][]{bhs+19}, any underlying tidal tails associated with this GC are sub-optimal for tracing the gravitational potential of our Galaxy. 
Moreover, this relatively small apocenter suggests that any underlying tidal tails from recent orbital wraps may not extend into the outer halo of the Milky Way, making them sub-optimal for probing the outer potential of our Galaxy or isolating any perturbations as being from dark subhalos.

\subsection{NGC 7492}
\label{sec:ngc7492}

\begin{figure*}[th!]
    \centering
    \includegraphics[width=\linewidth]{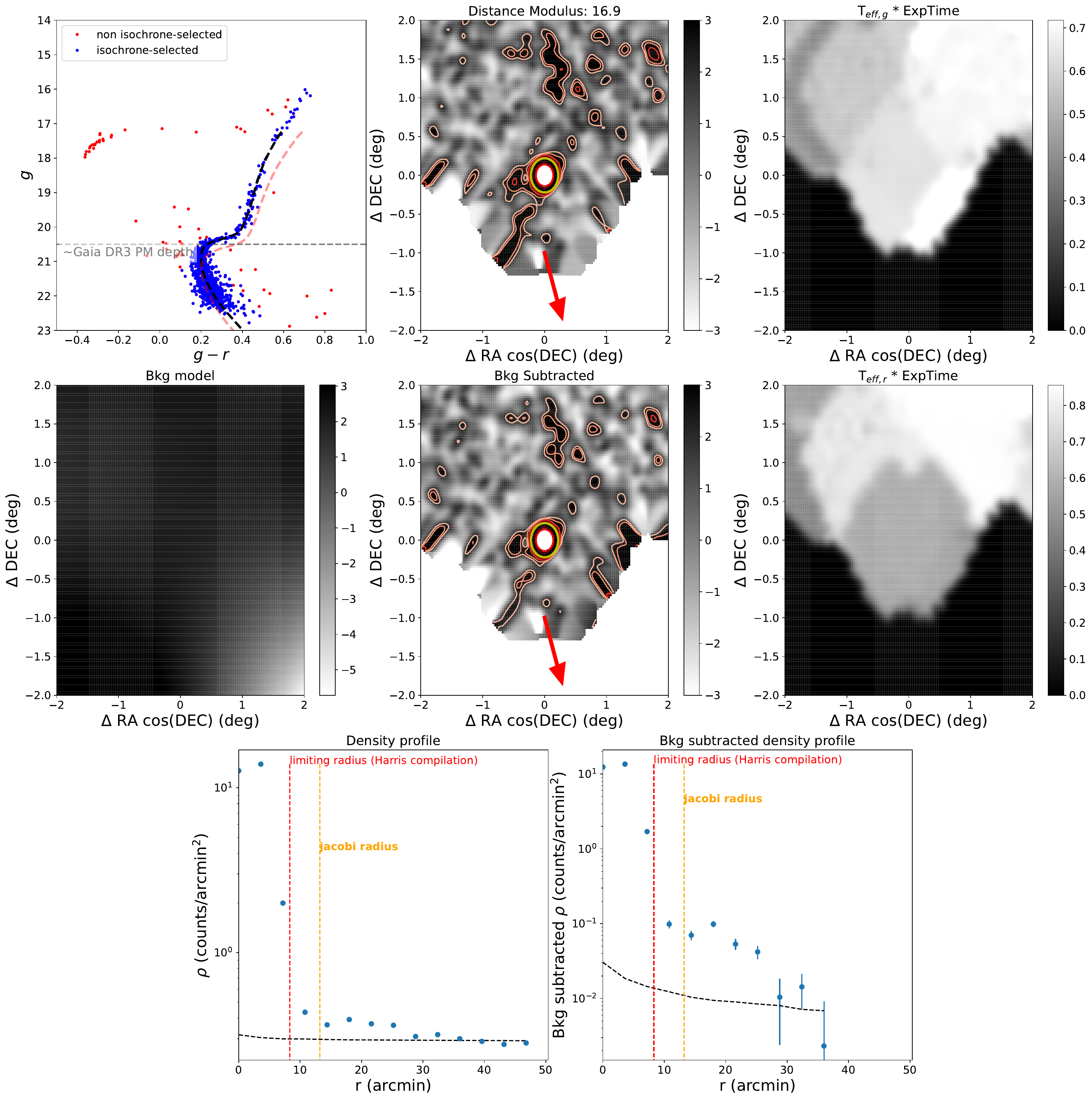}
\caption{Same as Figure~\ref{fig:pal5example}, but for NGC 7492. 
 A clear photometric excess is seen beyond the Jacobi radius of the GC (yellow circle).
 As with NGC 5897, this excess does not preferentially align with the proper motion of NGC 7492 (red arrow).}
\label{fig:ngc7492}
\end{figure*}

\begin{figure}[th!]
    \centering
    \includegraphics[width=0.4\textwidth]{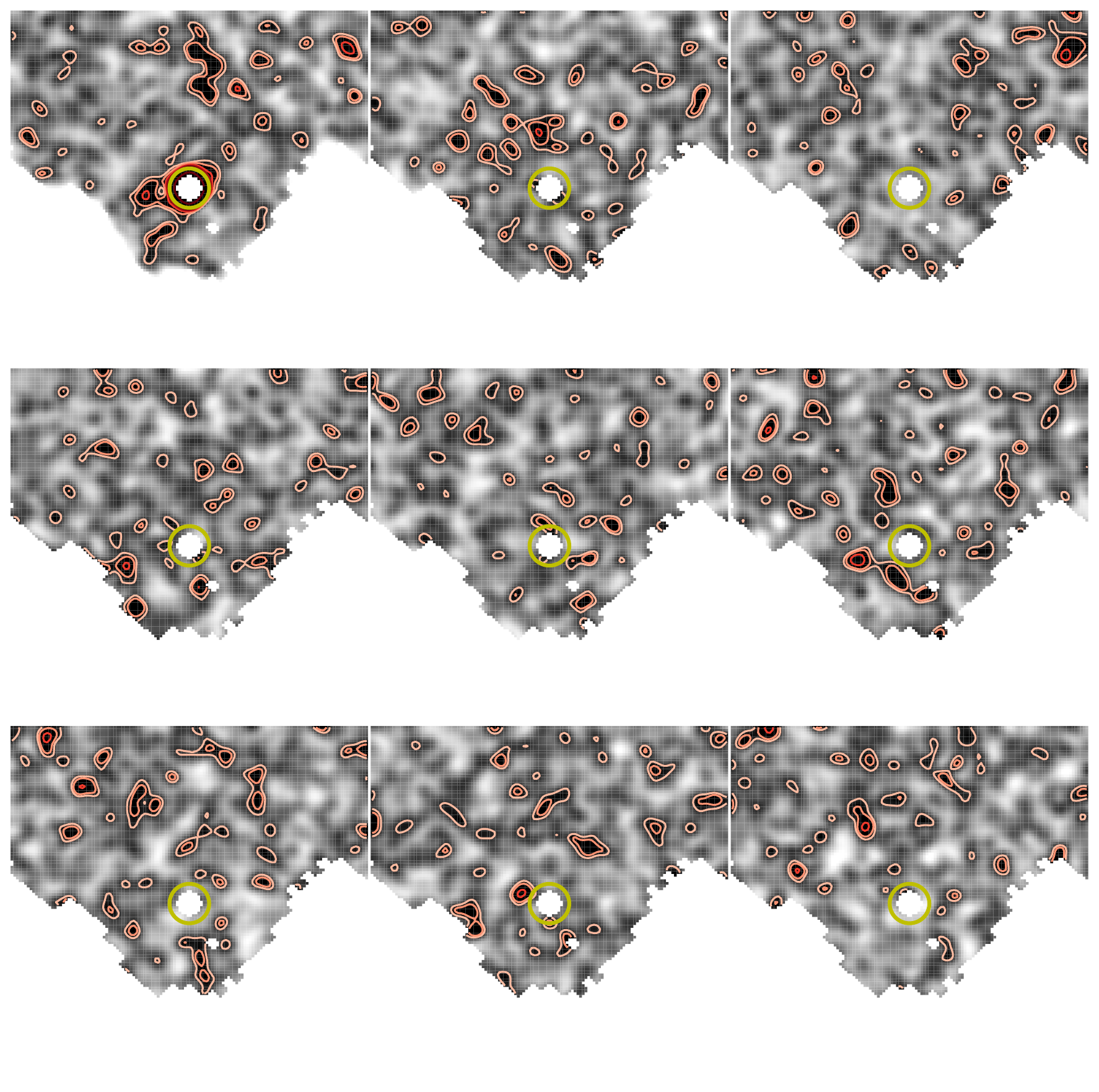}
    \includegraphics[width=0.4\textwidth]{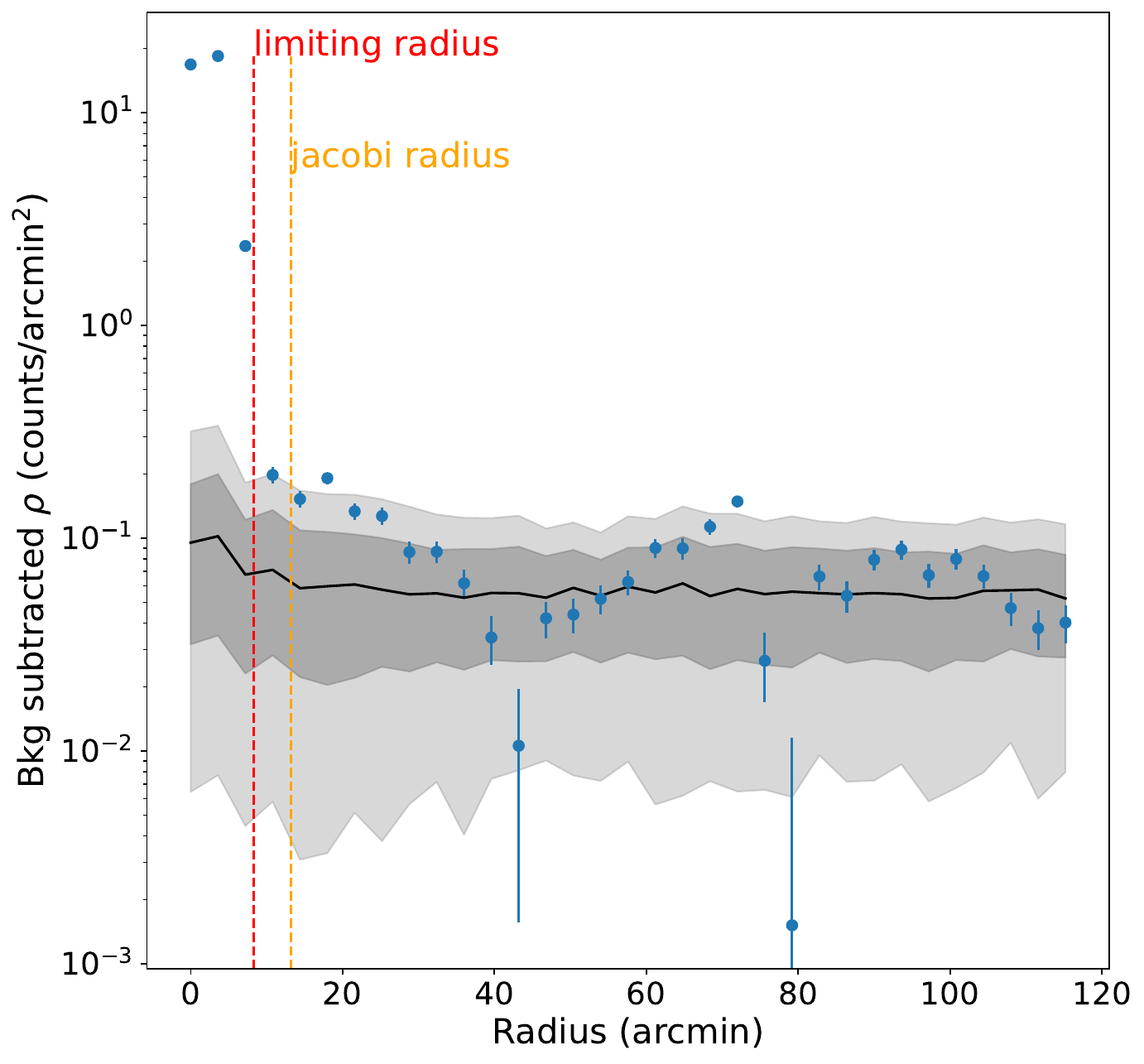}
\caption{Same as Figure~\ref{fig:ngc5897_mc}, but for NGC 7492. 
 The significance of the observed excess just beyond the Jacobi radius is 2.8\,$\sigma$, while being 1.9\,$\sigma$ at the Jacobi radius.}
\label{fig:ngc7492_mc}
\end{figure}

NGC 7492 ($\alpha = 347.11 \degree, \delta = -15.61 \degree$) is the other GC we discuss for which we find evidence of an extra-tidal envelope beyond $r_J$.
In the matched-filter density plot used to diagnose the presence of extra-tidal features (center panel of Figure~\ref{fig:ngc7492}), we see a photometric excess beyond $r_J$ (yellow circle).
Repeating the Monte Carlo analysis described in Section~\ref{sec:ngc5897_obs} for this cluster, we find a 2.8\,$\sigma$ excess slightly beyond $r_J$ (Figure~\ref{fig:ngc7492_mc}) with a $1.9\,\sigma$ excess at $r_J$, suggesting a tentative detection.
As with NGC 5897, the photometric excess does not preferentially lie in the same direction as the proper motion (red arrow in the center panel of Figure~\ref{fig:ngc7492}).
Unlike NGC 5897 however, NGC 7492 is on a more elliptical orbit (eccentricity $\approx 0.74$; \citealt{pwc19}) and approaching its apocenter (Galactocentric distance of 25.6, $r_{\textrm{apo}}\approx 28.2$; \citealt{bhs+19}), meaning it has not had a recent pericentric passage that can explain the photometric excess configuration.

This detection adds to a series of conflicting results in the literature about the existence of extra-tidal features around NGC 7492, and there is currently no consensus about whether or not NGC 7492 has extra-tidal features.
\citet{nbk+17} reported the discovery of tidal tails around NGC 7492 using PanSTARRS1 data.
However, follow-up studies with deeper photometry were unable to corroborate those results \citep{mcs+18a, mcs+18b, zmd+22}.
The discrepancy may originate in the fact that NGC 7492 lies in the path of Sagittarius (Sgr) dwarf galaxy stellar stream, which has significant overlap in the CMD with NGC 7492.
It is therefore crucial to separate the Sgr as much as possible and see if the extra-tidal densities persist.
Accordingly, we repeat our analysis using only stars that have proper motions in \textit{Gaia} DR3 \citep{gaia+23} that are consistent within 3$\sigma$ of membership to NGC 7492 ($\mu_\alpha$ = 0.80\,$\masyr$, $\mu_\delta$ = $-2.27\masyr$; \citealt{vasiliev19}).
This step ought to ensure a reasonable separation from the Sgr stream, since this region of the Sgr stream has a proper motion ($\mu_\alpha \sim -1.9\,\masyr$, $\mu_\delta \sim -3.3\masyr$; based on the Sgr stream membership catalog of \citealt{ramos+22}) that is well-separated from NGC 7492.
The same extra-tidal envelope seen in Figure~\ref{fig:ngc7492} persists when limiting to this sample, suggesting that this envelope is associated with NGC 7492 and not due to contamination by Sgr. 

\section{Dynamical and Stellar Population Modeling} \label{sec:modeling}

In this section, we present our procedure for modeling extra-tidal debris in the GCs in our sample.
The motivation for this exercise is to match our GC displaying new candidate extra-tidal features (NGC 5897) to predictions from dynamical modeling, and assess whether these extra-tidal features may indicate the presence of fainter tidal tails.
We also choose to dynamically model Pal 5 to test our method on a cluster with confirmed prominent tidal tails.
Additionally, we model NGC~5634, a GC with no extra-tidal features, to test whether a non-detection may still indicate fainter tidal tails.
We leverage the precise measurements of each GC's current 6D kinematic properties to simulate their past orbits around the Milky Way using the galactic dynamics package \texttt{gala} (Section~\ref{sec:simdebris}; \citealt{gala}).
We then generate synthetic stellar populations for each GC using the \texttt{SPISEA} package \citep{hll+20}, and apply observable magnitudes from these models to the aforementioned simulations (Section~\ref{sec:synth_pop}).
This allows us to deduce the surface brightness of extra-tidal features around these GCs, apply magnitude cuts to match the depth of our analysis in DELVE DR2, and then further forecast the detectability of streams around these GCs at the 10\,yr depth of LSST. 

We emphasize that answering whether these GCs may plausibly host fainter tails, and assessing their detectability with deeper data, is the primary objective of our dynamical modeling.
Consequently, we vary mass-loss parameters and foreground density until our simulations match our DELVE DR2 observations at their depth, and note that the detectability of extra-tidal features is a strong function of the amount of mass loss.
The physics of tidal stripping and GC evolution is governed by a number of factors and is an active area of research \citep[e.g.,][]{gwl+23, wkf+23, wrc+24, cvg+25}.
A full investigation of these processes is beyond the scope of this paper, but we aim to encapsulate their cumulative observational effect in our mass-loss and foreground adjustments  (e.g., paragraph 4 in Section~\ref{sec:simdebris}) to match observed data for the exploration in this section.

\subsection{Simulating Tidal Debris Formation}
\label{sec:simdebris}

In order to simulate the orbit of a GC around its host galaxy, we require its six-dimensional phase-space position and the potential in which it resides.
For most GCs, the present-day phase-space position is well constrained by numerous studies.
We use sky positions and proper motions from \citet{vb+21} and adopt distances and radial velocities from \citet{bh18}.

The Galactic potential, on the other hand, is an ongoing research topic with a number of proposed best-fitting potential models.
We choose the current best fit to observations (\texttt{MilkyWayPotential2022}) implemented in \texttt{gala} \citep{gala}.
\texttt{MilkyWayPotential2022} is a multi-component potential model including a spherical nucleus and bulge in the center of the Galaxy, an exponential disk consisting of a 3-component sum of Miyamoto-Nagai disks \citep{mn75}, and a spherical Navarro-Frenk-White (NFW) dark matter halo \citep{nfw+97}.
The disk model is fit to the \citet{ehrn19} rotation curve and the \citet{dhpj23} vertical structure.
The halo, bulge, and nucleus components are fit to recent mass measurements of the Milky Way (see \citealt{hpjd+22}).
We also include the LMC in our global potential, since it has been shown to affect the orbits of clusters and the dynamics of tidal tails \citep{dbl+19, kbl+19, slp+19, sed+21}.
Our LMC is an NFW potential object with a scale radius of $4$ kpc and a circular velocity of $90$ km/s at $5$ kpc. 
We place the Milky Way potential at the origin of our coordinate system at the present day and use values from \citet{vbe21} for the present day phase-space coordinates of the LMC, which incorporate the distance of 50\,kpc from \citet{fmg+01} and the line-of-sight velocity of 262.2\,km\,s$^{-1}$ from \citet{vk+14}.
Both objects are allowed to move within the space, although the shape of their potentials remains fixed.
While this is technically unrealistic since the LMC induces a dipole perturbation in the Milky Way halo \citep[\eg][]{w95, gbc+15, gbl+21, cng+21, lpe+23}, this likely has an insignificant effect on the scale of GC extra-tidal features ($\sim 2 \degree$) that we aim to explore.
We use this final combined MW-LMC potential for our GC orbit integrations.

Our simulation process contains two major components.
We first integrate the cluster orbit backward in time to a ``starting position".
Then, we populate the GC with particles and simulate their orbits to the present day to investigate the distribution of tidal debris in the vicinity of the GC.
Obtaining the starting position for a given GC is straightforward using \texttt{gala}'s \texttt{DirectNBody} class, which integrates orbits in a given potential using the Dormand-Prince method for solving ordinary differential equations \citep{dp80}.
We choose the backwards-integration duration to derive the ``starting position" as 3\,\Gyr.
While nearly every GC has been orbiting the Milky Way significantly longer than 3\,\Gyr, the early orbital history of these GCs is likely negligible when analyzing extra-tidal features in their immediate vicinity ($\sim 2 \degree$), since stars stripped $> 3 \Gyr$ ago tend to be further away from the progenitor \citep{c20} and long backward integrations increase the scope for errors \citep{db+22}.
Consequently, we adopt an integration time of 3\,\Gyr to offset the computational cost of longer integrations with minimal impact on our motivation of assessing debris in the vicinity of GCs.
We note that since the GC velocity uncertainties are small \citep[$\mu_{\textrm{RA,err}}, \,\mu_\textrm{Dec,err} \leq 0.03 \masyr$, $v_{r,\textrm{err}} \leq 0.5 \kms$;][]{vb+21,bh18}, we do not propagate the effects of uncertainties in the astrometry in this exercise, which could lead to varied disruption scenarios.

The initial 3\,\Gyr backward integration provides our ``starting" 6D phase space position of the GC, at which point we add individual particles to the GC and simulate tidal debris by integrating the orbits of these particles forward to the present day.
We populate the GC with particles by modeling its stellar distribution as a Plummer profile, and its mass distribution with a Plummer potential \citep{plummer11}.
To define these quantities, we require a plummer scale radius ($r_p$) and a total mass.
We adopt $r_p$ for our GCs of interest from the compilation of \citet{vb+21}.
We require a bespoke approximation to obtain the total initial GC mass for our simulation (at 3\,Gyr ago), since prior studies largely provide GC mass estimates at the present day or initial masses before any mass loss \citep{bh18}.
For our initial mass calculation, we perform an initial ``test-run" of the orbit to estimate the fraction of mass lost over the 3\,\Gyr orbit to the present day, by initializing a Plummer potential with a total mass equal to the present day GC mass, populating it with 1000 test particles, and integrating their orbits within the MW-LMC-GC combined potential.
Every 50 \Myr, we update the mass of the GC potential by establishing a boundary outside of which stars are labeled as ``lost'', in line with a similar method used by \citet{im+24} where $10\,r_p$ is used.
For each lost particle, we lower the mass of the GC potential by 0.1\% of the starting mass and iterate until the present day, to estimate total mass-loss over the 3\,\Gyr period.
Over this period for NGC 5897 and NGC 5634, we find that this exercise suggests mass losses of 25\,\% and 10\,\%, respectively.
For Pal 5, we artificially extend the mass-loss boundary to $50\,r_p$ to force a match to observations (see Section~\ref{sec:modeling_analysis}; Figure~\ref{fig:pal5_sim}), as the fiducial value of $10\,r_p$ leads to significant over-disruption.
Since the primary aim of our exercise is to match observations and our mass-loss modeling is fairly simple, we find it motivated to tune this multiple of $r_p$ to force a better match with observations.
We then multiply the current GC mass in \citet{bh18} by the inverse of the fractional mass loss from this exercise to derive a GC mass 3 Gyr ago for orbit integration. 

Then, we perform our full simulation of the GC by taking its initial conditions, updating its potential to reflect the initial mass estimate, and populating it with non-interacting test particles to approximate its number of stars.
Specifically, we initialize N=M$_*$/0.43\,M$_\odot$ test particles following a Plummer density profile in the GC potential, following the assumption of an average mass of 0.43\,M$_\odot$ per star \citep{bhd+23}.
We then calculate the location of each of these particles at the present day by integrating forward for 3\,\Gyr, following the same mass-loss prescription outlined in the previous paragraph, to trace particles populating extra-tidal features in the vicinity of the GC.
We note that in our analysis, we have ignored explicitly treating aspects of both GC evolution and physics (e.g., three-body interaction, mass-segregation) that affect the population of stars in tidal tails or extra-tidal envelopes. 
This is largely due to computational feasibility (e.g., straightforward parallelization of orbit integrations) and the scope of our initial motivation, which is to assess whether the GCs that we detected with extra-tidal envelopes may host faint underlying tidal tails, and to qualitatively assess to what extent deeper LSST photometry may increase sensitivity.
Due to this observational motivation, we perform a check to ensure that the simulated envelopes roughly match observations at the DELVE DR2 depth (see Section~\ref{sec:modeling_analysis}) as a check on our mass-loss prescription, but note that a detailed analysis of the sources of GC mass-loss are beyond the scope of this work. 

\subsection{Adding Synthetic Models of Stellar Populations} \label{sec:synth_pop}

We then generate synthetic simple stellar populations for each GC using \texttt{SPISEA} \citep{hll+20}\footnote{https://spisea.readthedocs.io/en/latest/} and assign magnitudes and colors to each particle in the simulation in Section~\ref{sec:simdebris}.
Briefly, \texttt{SPISEA} is a python package that is used to generate stellar populations, with inputs for a number of parameters including the initial mass function (IMF), metallicity, and age. 
We generate stellar populations in \texttt{SPISEA} matching the number of particles in the \texttt{gala} simulations, assuming  default parameters from the \texttt{SPISEA} guide (e.g., Kroupa IMF, \citealt{k01}; MIST isochones, \citealt{d+16}, \citealt{cdc+16}, \citealt{pbd+11,pca+13,pms+15,psb+18}; multiplicity following \citealt{ldg+13}), but fixing the age and metallicity to match those of the particular GC \citep{h+96}.
The color/magnitude for each star in the synthetic stellar population outputted by \texttt{SPISEA} is stored in the DECam $g,r$ filter system and randomly assigned to each particle in the simulation.
The magnitudes assigned to each particle are then converted to apparent magnitudes by computing the heliocentric distances in the \texttt{gala} simulation.
After these steps, we have a clean framework to filter particles in the \texttt{gala} simulation to match the parameters of our observational search (e.g., photometric depth).

We note that, to facilitate a consistent comparison between the simulation and our observational search, it is essential to use a foreground model of the Milky Way to populate the simulation.
Accordingly, to replicate the parameters of our search in Section~\ref{sec:obs_analysis}, we populate a 2$^\circ$ region around each GC with a subsample of the catalog output from LSSTsim DR2 \citep{dpm+22}.
LSSTsim is a model of the stellar population of the Milky Way based on the TRILEGAL code \citep{ggh+05} down to $r=27.5$, specifically designed to forecast for LSST.
We note that single stars in LSSTsim DR2 are simulated using evolutionary tracks following \citet{mgb+17}, which uses \texttt{PARSEC} v1.2S \citep{bmg+12} and \texttt{COLIBRI} PR16 \citep{mbn+13, rmg+16}, and the detectability exploration in our simulation is sensitive to faint-end stellar population assumptions in these tracks.
LSSTsim DR2 is hosted on NOIRLab's datalab servers\footnote{https://datalab.noirlab.edu/lsst\_sim/index.php}, facilitating queries to this catalog.
We accordingly query LSSTsim DR2 for all stars within the aforementioned 2$^\circ$ region around each GC, and convert the Rubin/LSST $g,r$ magnitudes to DECam $g,r$ using the following equations \citep{dmtn277}\footnote{https://github.com/lsst/throughputs/}:
\begin{equation}
\begin{split}
g_{\textrm{DES}} = g_{\textrm{LSST}} + 0.0067 - 0.0341\times(g-i)_{\textrm{LSST}} + &\\
0.0295\times(g-i)_{\textrm{LSST}}^2 - 0.0141\times(g-i)_{\textrm{LSST}}^3
\end{split}
\end{equation}
\begin{equation}
\begin{split}
r_{\textrm{DES}} = r_{\textrm{LSST}} + 0.0435 - 0.1830\times(g-i)_{\textrm{LSST}} + &\\0.0519\times(g-i)_{\textrm{LSST}}^2 - 0.0131\times(g-i)_{\textrm{LSST}}^3
\end{split}
\end{equation}
We subsample a portion of this catalog (15\% -- 20\%; see Sections~\ref{sec:pal5} to~\ref{sec:ngc5634}) to visually reproduce the observed density plots, and then append this subsample to the simulated magnitudes of the particles from the simulations of GC disruption to create a final catalog of simulated GC stars and Milky Way foreground.

We then pass this final catalog through the search algorithm we describe in Section~\ref{sec:obs_analysis} to assess whether our code also picks up extra-tidal envelopes or streams around GCs in the simulations, with two modifications. 
First, we exclude the RGB portion of the CMD in our selection (e.g., as in Figure~\ref{fig:pal5_sim} due to misalignment between the simulated and observed CMDs in this regime. 
Second, we omit the proper motion selection criteria to exclude obvious non-members in the observational analysis. 
With these modifications, we run this test at the depth of DELVE DR2 for each GC to ascertain whether extra-tidal envelopes show up at these depths in the simulations.
As a final empirical adjustment, we shifted the $g_{DES}$, $r_{DES}$ magnitudes from SPISEA slightly (0.1\,mag in $g$, 0.25\,mag to 0.32\,mag in $r$) to force a match to the observed CMDs for each GC, and random scatter was added to the simulated magnitudes to reproduce observational photometric uncertainties, following the functional forms from \texttt{ugali} \citep{bdb+15}\footnote{https://github.com/DarkEnergySurvey/ugali/}: 
\begin{equation}
\begin{split}
\sigma_{g} = \exp(-0.978\times(g_{10\sigma\,\textrm{lim}} - g) -2.322) + 0.015
\end{split}
\end{equation}
\begin{equation}
\begin{split}
\sigma_{r} = \exp(-0.979\times(r_{10\sigma\,\textrm{lim}} - r) -2.198) + 0.013
\end{split}
\end{equation}
with 10$\sigma$ limiting magnitudes taken from the DELVE DR2 release paper ($g = 23.5$, $r = 23.1$; \citealt{dfa+22}).
We then repeat the same procedure at the forecasted 10\,yr depth of LSST ($g = 26.9, r = 26.9$; \citealt{bij+22})\footnote{https://www.lsst.org/scientists/keynumbers} to assess whether these envelopes may indicate the presence of observable tidal tails in upcoming deeper photometric datasets. 
We show the case for the LSST 10\,yr depth to investigate what the deepest future dataset may provide, with the understanding that intermediate photometric datasets would lie somewhere between this and DELVE DR2.
We opt for these general forecasted photometric depths as opposed to tuning per pointing, to more generally illustrate the gain from deeper photometry relative to DELVE DR2.
We note that a more realistic depth map for the 10\,yr LSST depth evaluated at the specific locations of our three clusters reveals depths within $0.2\magn$ of this chosen $26.9\magn$.

\section{Modeling Analysis \& Application to Observational Results}\label{sec:modeling_analysis}

\begin{figure*}[th!]
    \centering
    \includegraphics[width=0.9\linewidth]{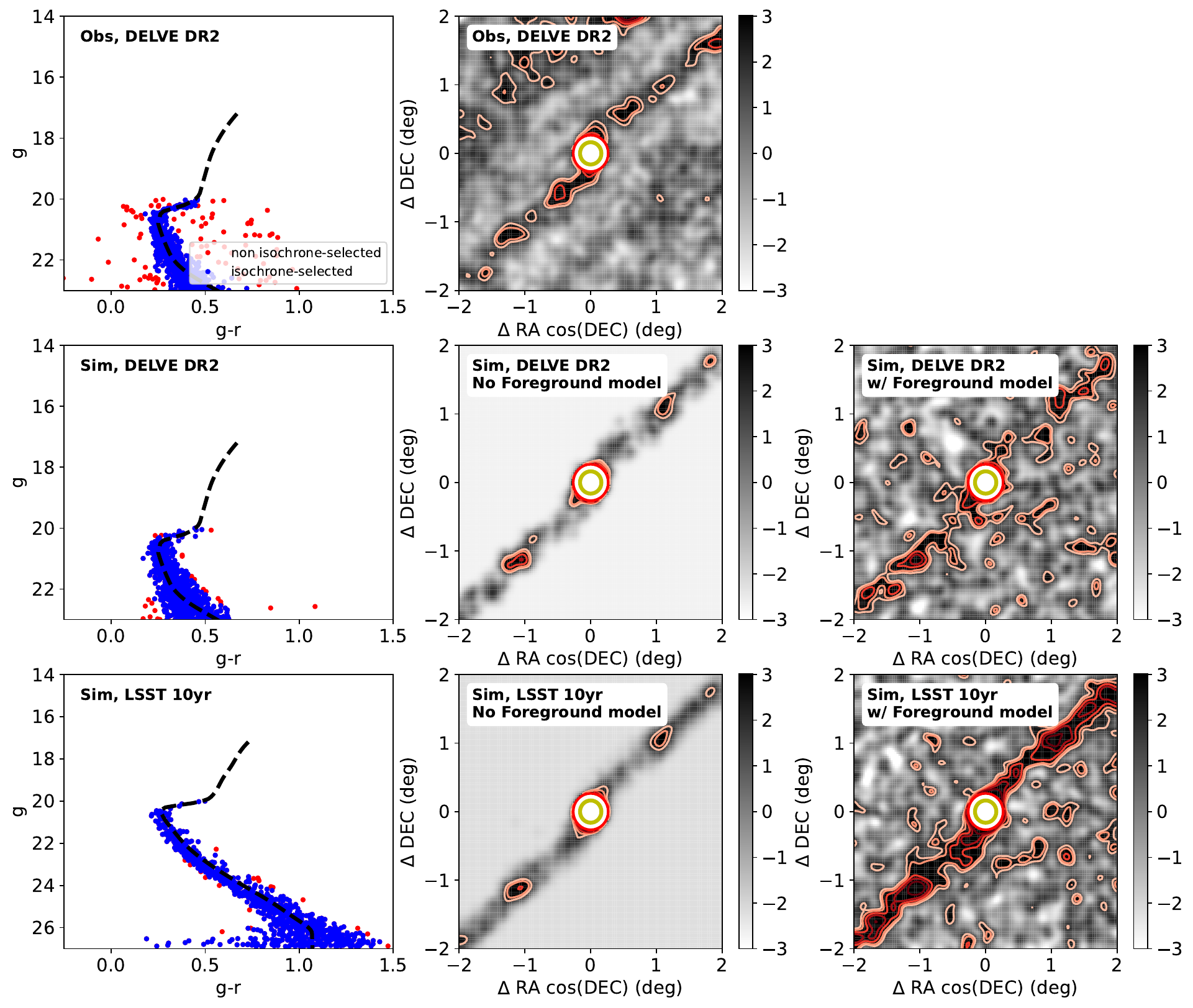}
\caption{A comparison of the observed Pal 5 stream in DELVE DR2 (top panels) with a simulation of the stream down to the DELVE DR2 depth (middle panels) and the LSST 10\,yr depth (bottom panels). 
The left panels show the observed CMD (top), the simulated CMD from SPISEA \citep{hll+20} with a DELVE DR2 magnitude error function (middle), and an error function for the LSST 10\,yr depth (bottom).
Stars selected along the isochrone for Pal 5 are shown in blue; note that stars along the red giant branch are excluded in the selection due to misalignment in the simulated and observed CMDs in that regime.
The middle simulated column shows the isochrone selection applied to the simulation of Pal 5 with no Milky Way foreground, and the right simulated panels show the isochrone selection in the presence of a Milky Way foreground model populated by a fraction of stars in LSSTsim DR2 \citep{dpm+22} to force a rough visual match with observations at DELVE DR2 depth ($g=23.25$, see Section~\ref{sec:modeling_analysis}).
As expected, the stream becomes significantly more prominent at the forecasted full depth of LSST.}
\label{fig:pal5_sim}
\end{figure*}

In this section, we present the results of our simulations of three GCs (Pal 5, NGC 5897, NGC 5634).
Pal 5 was chosen as a fiducial example of a GC with a known stream that we detect in the DELVE DR2, in order to demonstrate that the simulations can also produce tidal tails that become more significant at larger depths. 
We simulate NGC 5897 to assess whether the detected extra-tidal envelope is consistent with the presence of fainter extra-tidal tails that are either obscured by foreground or could be detectable with the full 10\,yr LSST depth.
Finally, we simulate NGC 5634 to provide an example of a non-detection and test whether this implies a lack of even fainter tidal tails.
We do note that the mass-loss prescription and foreground level in the simulations have been tuned to roughly reproduce the observations in DELVE DR2 (see Section~\ref{sec:pal5}, Figures~\ref{fig:pal5_sim},~\ref{fig:ngc5897_sim},~\ref{fig:ngc5634_sim}), before extending to the LSST depth.
Consequently, these exercises should be viewed as toy models to help interpret the extra-tidal envelopes around NGC~5897 and qualitatively assess detectability at deeper depths, as opposed to exact simulations of the physics of GC mass loss. 

\subsection{Pal 5}
\label{sec:pal5}

In Figure~\ref{fig:pal5_sim}, we show the results of our simulation on Pal 5.
The mass-loss prescription and foreground contamination in the simulation were tuned to visually match the significance of the over-densities at DELVE DR2 depth. 
Specifically for Pal 5, this required setting the mass-loss boundary 50\,$r_p$ (see Paragraph~4 in Section~\ref{sec:simdebris}), notably higher than the standard value in the literature that was used in the simulations for NGC 5897 and NGC 5634. 
The foreground contamination was set by including 15\% of the LSSTsim DR2 catalog.
The depth in both the simulation and observations were set to $g=23.2$ to minimize systematics in the observed data from depth variations.

At a high-level, we recover evidence of tidal tails in the simulation at the DELVE DR2 depth (middle panel) and with increased prominence at the full LSST 10\,yr depth (bottom middle panel) which is considerably fainter ($\sim$3\,mag).
The detection of the stream persists in the presence of Milky Way foreground in the simulation (right panels). 
This indicates that additional work on Pal 5 with LSST could be fruitful, as it will offer the opportunity to explore the stream with significantly more fidelity near its progenitor. 
The immediate vicinity of GCs undergoing tidal stripping is thought to be potentially under-dense \citep{bnc+21}, and increased sensitivity here might be instructive on the physics of tidal stripping processes.

For the purposes of this study, the ability of the simulation to produce a realistic stream around Pal 5 encourages us to use these simulations as a framework to assess the extra-tidal envelopes around NGC 5897 and NGC 5634.
Qualitatively, this exercise with Pal 5 demonstrates that the simulation can generate underlying streams around a GC under the aforementioned assumptions of mass-loss and foreground contamination, that become more prominent with deeper observations from stellar population modeling.

\subsection{NGC 5897}
\label{sec:ngc5897}

\begin{figure*}[th!]
    \centering
    \includegraphics[width=0.9\linewidth]{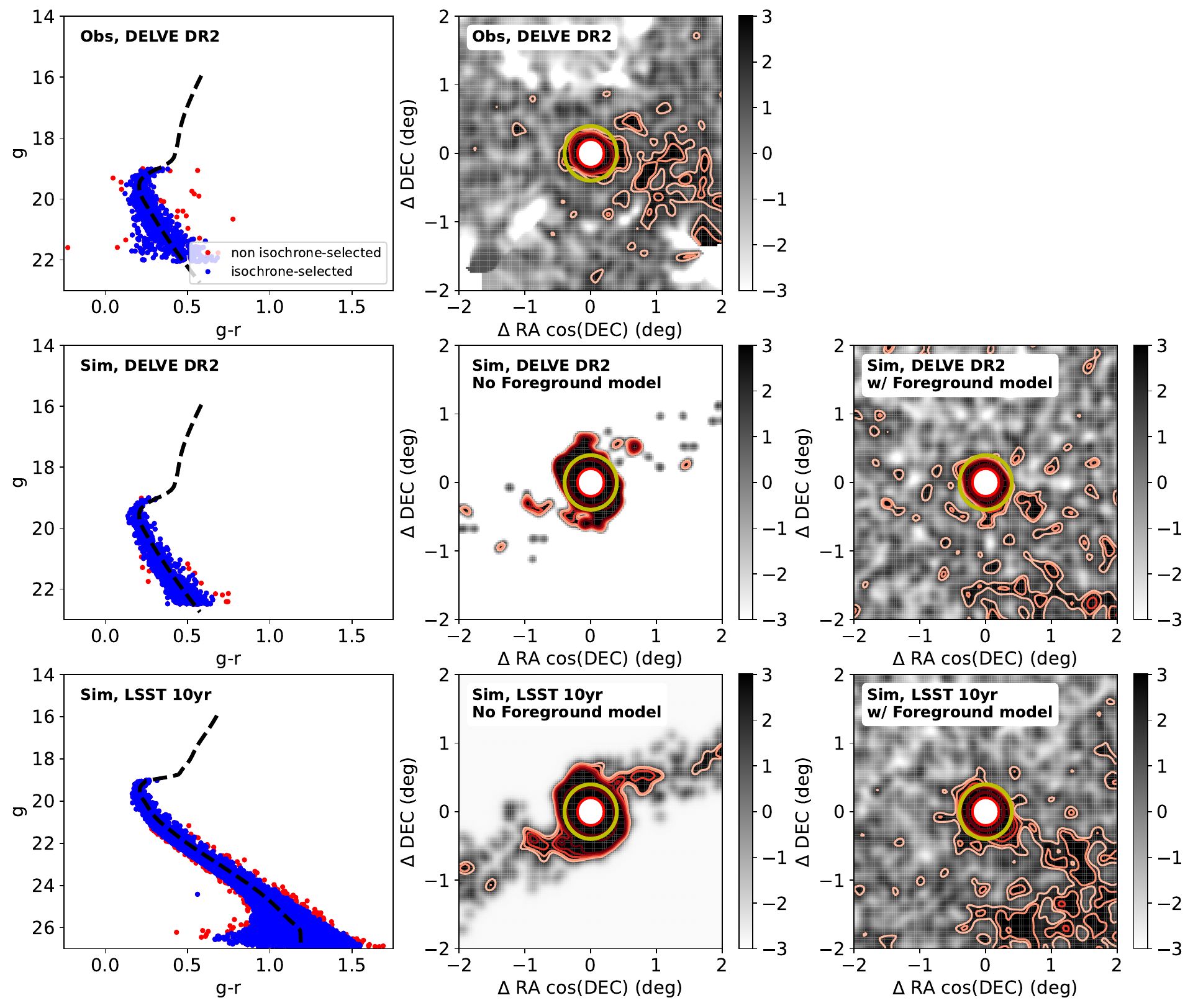}
\caption{Same as Figure~\ref{fig:pal5_sim}, but for NGC5897. 
At the depth of DELVE DR2, our model suggests that the underlying tidal tail of this GC would only barely be detectable, even in the presence of no Milky Way foreground (middle panel).
At the full LSST depth, the same model suggests that the stream ought to be detectable in the presence of no Milky Way foreground. 
The addition of a Milky Way foreground model in both cases masks the underlying stream and leaves only the extra-tidal envelope detectable. 
}
\label{fig:ngc5897_sim}
\end{figure*}

Our simulation for NGC 5897, which is the GC that hosts our most promising detection of new extra-tidal features, is shown in Figure~\ref{fig:ngc5897_sim}.
The foreground level was set to 20\,\% of the total in LSSTsim DR2 to approximately visually match the simulated GC at DELVE DR2 depth (middle right panel of Figure~\ref{fig:ngc5897_sim}) to the observed DELVE DR2 density plot (top middle panel), in the same manner as in Section~\ref{sec:pal5}.
These parameters also result in the simulation having a 3.3\,$\sigma$ statistically significant extra-tidal over-density at $r_J$ at DELVE DR2 depth.
Notably, the presence of an extra-tidal envelope appears in the same manner in both simulation and observations.

Extrapolating to the LSST 10\,yr depth is an illustrative exercise for NGC 5897, as it suggests that an underlying faint stellar stream in the vicinity of the GC is in principle detectable if there were no foreground (bottom middle panel of Figure~\ref{fig:ngc5897_sim}), but would likely be obscured by Milky Way foreground (bottom right panel). 
However, at DELVE DR2 depth, our simulation suggests that the underlying stellar stream would only barely be detectable, even if we fully removed foreground contamination (middle panel). 
Overall, the contrast of the GC relative to the background becomes stronger at increased depth and the detection of an envelope becomes more significant ($\sim10.5\,\sigma$ detection at $r_J$).
However, we do not recover extended tidal tails from photometry alone, suggesting that additional membership discrimination from foreground (e.g., using LSST proper motions, LSST $u$-band metallicities), or searches over larger fields of view will be important in studying these systems.
Scientifically, our simulation demonstrates that NGC 5897's extra-tidal envelope (as seen in the observations; simulated in right panels in Figure~\ref{fig:ngc5897_sim}) is consistent with a faint underlying stream that is observationally obscured, oriented perpendicular to the envelope, and aligned with its proper motion (see models with no foreground in Figure~\ref{fig:ngc5897_sim}).

\subsection{NGC 5634}
\label{sec:ngc5634}
\begin{figure*}[th!]
    \centering
    \includegraphics[width=0.9\linewidth]{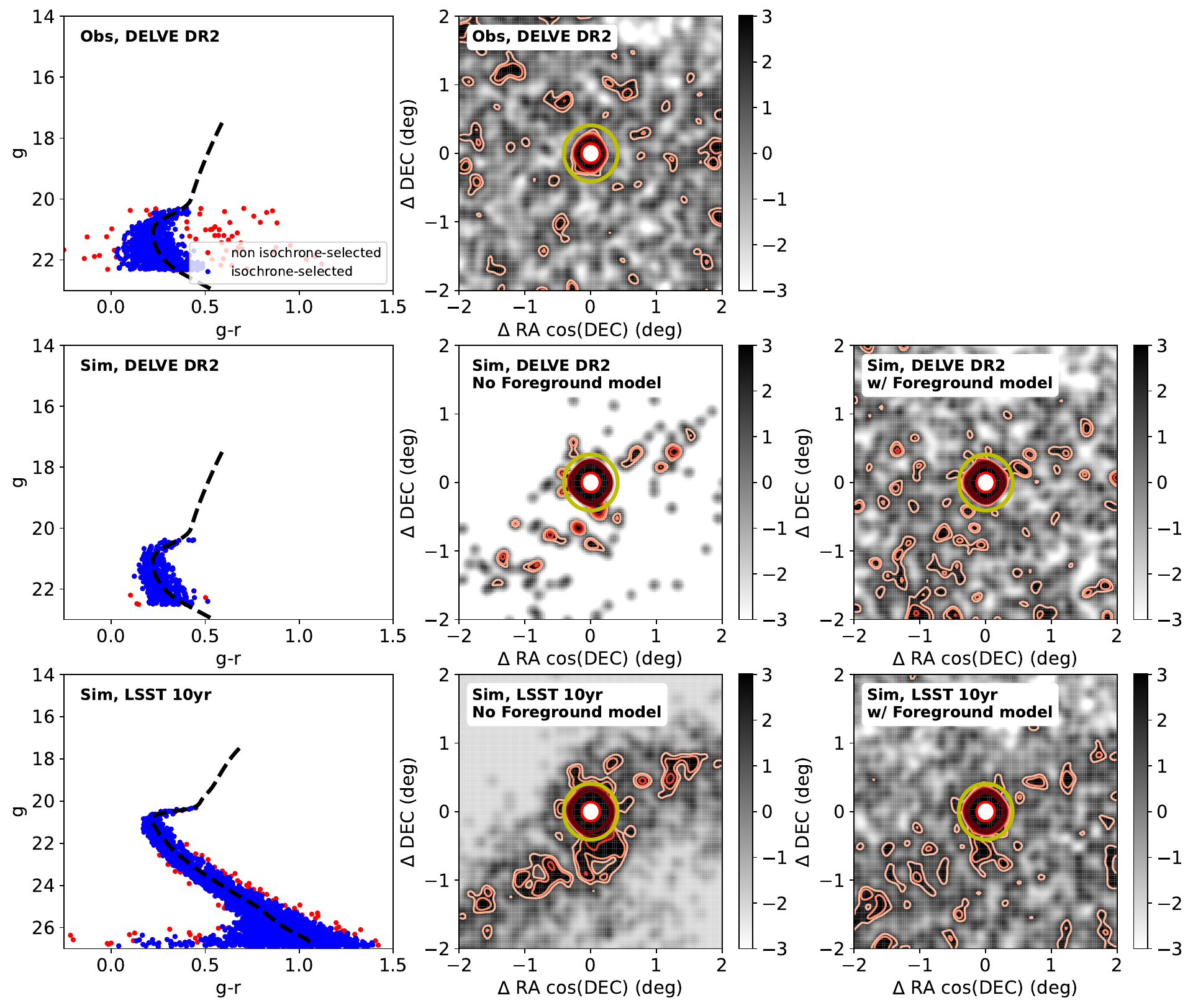}
\caption{Same as Figure~\ref{fig:pal5_sim}, but for NGC5634. 
As with NGC5897 (Figure~\ref{fig:ngc5897_sim}), at the depth of DELVE DR2 our model suggests that the underlying tidal tail of this GC is marginally detectable in the presence of no Milky Way foreground (middle panel).
At the full LSST depth, the same model suggests that the stream ought to be detectable in the presence of no foreground and potentially even with the addition of Milky Way foreground (bottom right panel); although, likely only with updated search parameters (e.g., a larger smoothing kernel, wider field-of-view).}
\label{fig:ngc5634_sim}
\end{figure*}

The simulation outcome for NGC 5634, shown in Figure~\ref{fig:ngc5634_sim}, offers additional perspective.
The foreground level is set to 30\% of LSSTsim DR2, and the additional depth of LSST improves the contrast of the GC relative to the foreground.
In the simulation, the cluster does develop extended tidal tails which are much wider in on-sky projection than the cluster itself as well as wider than the Pal 5 and NGC 5897 tails.
We theorize that this is likely because NGC 5634 is on a more eccentric orbit ($e\sim0.7$; \citealt{pwc19}) and recently passed its apocenter.
Despite these tails in the simulation, there is no clear detection of statistical significance at $r_J$ after including the foreground.
However, it is notable that the majority of the densest localities in the region lie along the track of the simulated stream at the depth of LSST.
While this may not be enough for a conclusive detection with LSST with our choice of search parameters, it is certainly promising and suggests that a search for its tails would be worthwhile with different parameters (e.g., a larger smoothing kernel, field-of-view).
Therefore, our simulation forecasts that the additional depth provided by LSST may principally create a difference in confirming observable extra-tidal features of NGC 5634, even without additional tools for foreground removal.
Scientifically, this exercise with NGC 5634 suggests that a lack of extra-tidal features at the DELVE DR2 depth does not imply the lack of a faint underlying stellar stream in this case, and that the stream becomes meaningfully more sampled with deeper photometry (middle column of Figure~\ref{fig:ngc5634_sim}).

\subsection{LSST Prospects \& Addressing Foreground Contamination}
\label{sec:lsstprospects}
The three GCs that were simulated span different outcomes when assessing the gain at LSST depth for detecting faint extra-tidal features around these objects.
Our results for Pal 5 indicate that LSST will improve the contrast of its already clearly detected tidal tails relative to the background, and offer a clearer picture of the area immediately surrounding the GC.
On the other hand, NGC 5897 does not show a qualitative difference in the detection of extra-tidal features between DELVE DR2 and LSST implementing a foreground model.
Lastly, the NGC~5634 simulation suggests that LSST may detect weak evidence for overdensities along the forecasted debris track.
We also note that for all three GCs, the reported exercise is closer to an upper limit on what can be detected in the presence of foreground since our model does not include galaxies and star/galaxy separation will be an additional confounder in the true data.

While these results offer differing outcomes for studying extra-tidal features around GCs with LSST, we list a few concrete points here.
At a scientific level, our simulations show that it is plausible for our detected extra-tidal envelopes to imply the existence of faint underlying tidal tails.
This is largely consistent with theoretical tidal stripping models, but interestingly, it has so far not been obvious from observations as many GCs exhibit tidal envelopes without obvious tails \citep[e.g.,][]{zmd+22}. 
Notably, even at DELVE DR2 depth with no foreground contamination, the central panels of Figures~\ref{fig:ngc5897_sim} and \ref{fig:ngc5634_sim} indicate that it would be difficult to find evidence for tidal tails around NGC 5897.
At LSST depth, however, the tails do become clear when foreground contamination is removed, but again become challenging to detect when foreground is incorporated.

This outcome highlights the necessity for additional methods of separating member and non-member stars in studying these faint features in the immediate vicinity of GCs (and faint, low metallicity features in general).
One promising prospect is to use proper motions, either from LSST, Euclid \citep{Euclid:24}, or the Nancy Grace Roman Space Telescope \citep{sbc+19}.
Another option is to use photometric metallicities from the LSST u-band filter \citep{lsst09}, the new Mapping the Ancient Galaxy in CaHK survey (MAGIC; Chiti et al., in prep; \citealt{bcl+25}) on DECam, the Pristine Survey \citep{smy+17}\footnote{We note that as our paper was nearing submission, \citet{kik+25} was posted on the arXiv using photometric metallicities to analyze GC extra-tidal features.}, or other similar undertakings.
Finally, the success that spectroscopic surveys like the Southern Stellar Stream Spectroscopic Survey \citep[$S^5$;][]{lkz+19} have had in characterizing streams has led to hope that ongoing/future spectrographs such as the Dark Energy Spectroscopic Instrument \citep[DESI;][]{desi22, cka23}, the 4-metre Multi-Object Spectroscopic Telescope \citep[4MOST;][]{dab+19}, or the WHT Enhanced Area Velocity Explorer \citep[WEAVE;][]{jtd+24} can also help detect underlying tidal tails.
Our work forecasts that these are promising avenues for the next decade of study on GC extra-tidal features and tails, with potential for uncovering 
more of these faint features.

\subsection{Predicting GC disruption} \label{sec:predict_stripping}

Looking ahead, we can also estimate which GCs should be experiencing strong tidal disruption effects in the Milky Way. 
We investigate this by comparing the Jacobi radius of each GC with its half-light radius and King limiting radius, following a similar approach as \cite{pel+22}. 
We use a slightly different set of parameters for these simulations, in which the Milky Way potential is from \cite{m+17}, and the LMC is modeled as a Hernquist profile \citep{h+90} with a mass of $1.38\pm0.255\times10^{11} M_\odot$ \citep{dbl+19} and a scale radius such that the mass enclosed within 8.7 kpc equals $1.7\times10^{10} M_\odot$ \citep{vdmk+14}. For the LMC's present-day phase-space coordinates, we use proper motions from \cite{kvb+13}, radial velocity from \cite{vdmah+02}, and distance from \cite{pgg+19}. We take the present-day phase-space coordinates of each GC from \cite{vb+21,bv+21}. We sample the orbit of each GC 100 times, accounting for the uncertainties in its present-day position and velocity, the Milky Way potential \citep[using the posterior chains of][]{m+17}, and the LMC's mass and present-day position and velocity. For each orbit, we compute the Jacobi radius at pericenter and apocenter using Equation~\ref{eq:tidal_radius}. 

\begin{figure*}[th!]
    \centering
    \includegraphics[width=\linewidth]{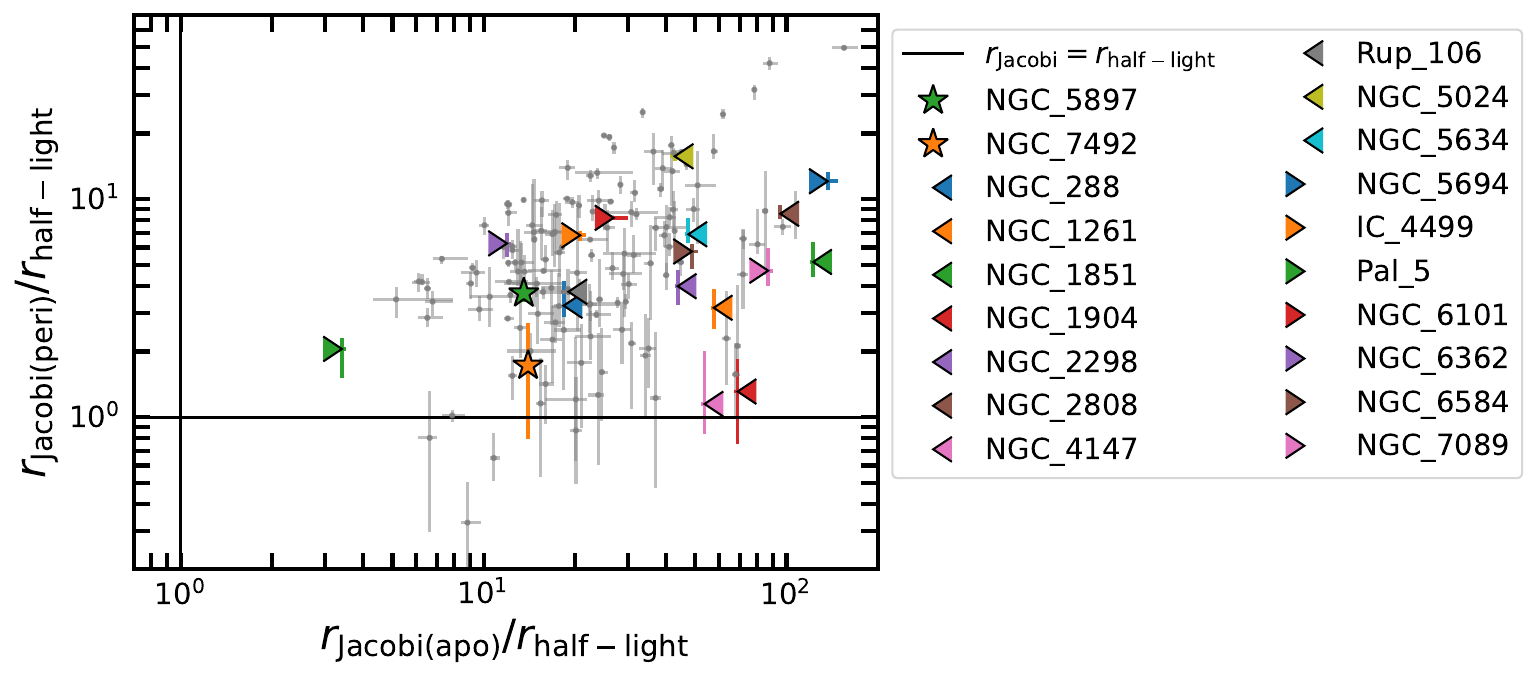}
\caption{Comparison of the Jacobi radius at apocenter and pericenter with each GC's half-light radius. For systems where these are comparable, the GC should experience strong tidal disruption. We show the tidal radius at pericenter and apocenter to highlight the range of tides along the orbit. Palomar 5, the first GC with a detected stellar stream, stands out as having the smallest ratio of its Jacobi radius at apocenter to its half-light radius.
The GCs studied in this work are shown as colored markers, and the entire GC population is shown in grey.}
\label{fig:rtidal_vs_rhalf}
\end{figure*}

\begin{figure*}[th!]
    \centering
    \includegraphics[width=\linewidth]{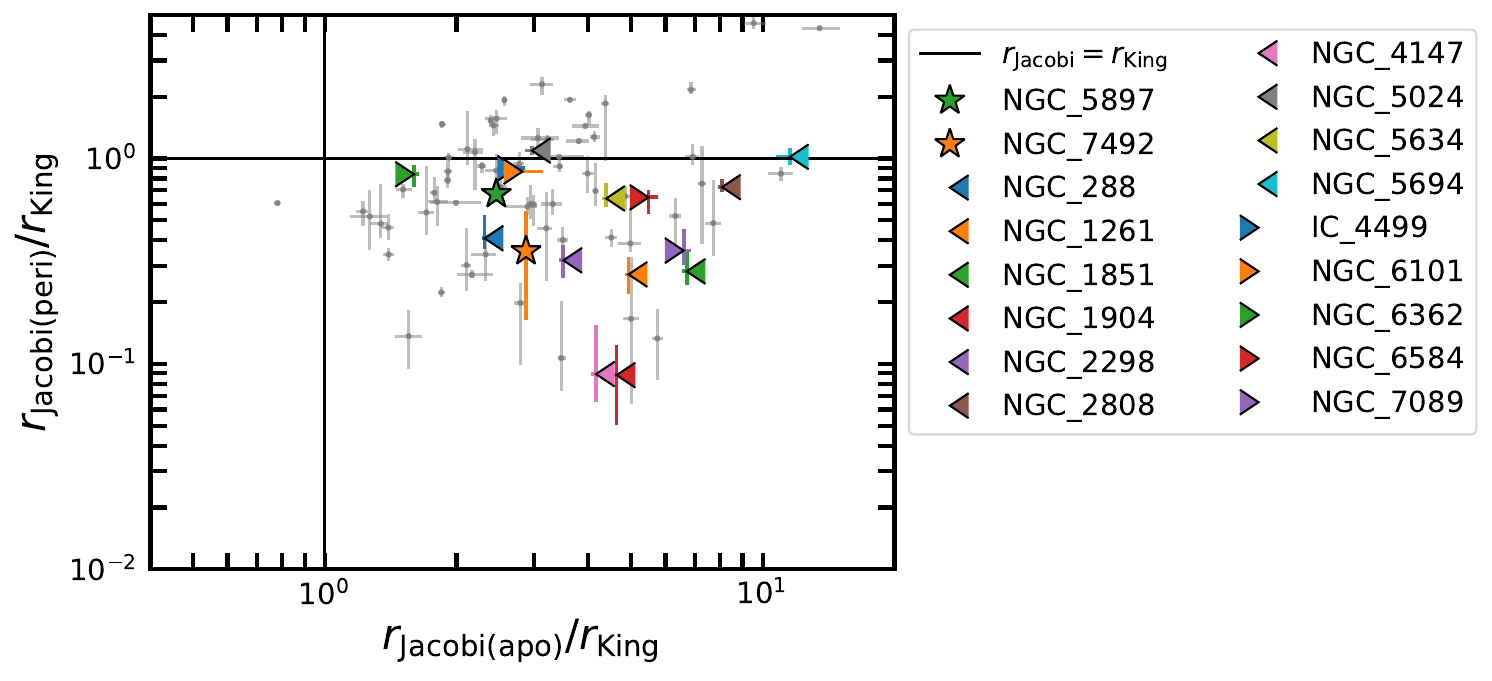}
\caption{Comparison of the Jacobi radius at apocenter and pericenter with each GC's King limiting radius. All but one of the systems studied in this work have a Jacobi radius at pericenter that is smaller than its limiting radius. }
\label{fig:rtidal_vs_rKing}
\end{figure*}

We compare these Jacobi radii with the half-light radius and King limiting radius of each GC in Figures~\ref{fig:rtidal_vs_rhalf} and \ref{fig:rtidal_vs_rKing}, respectively.
To ensure homogeneously derived values of structural parameters, we adopt half-light radii from \citet{bh18, bhs+19}\footnote{Accessed online at:\\
people.smp.uq.edu.au/HolgerBaumgardt/globular/parameter.html} and limiting radii from \citet{dgb+19}.
The Jacobi radii in these plots are listed in Table~\ref{tab:jacobi}. 
If the Jacobi radius at pericenter is comparable to the half-light radius, strong tidal disruption is possible since a large fraction of the stars can be stripped at each pericenter. 
In contrast, if the Jacobi radius is smaller than the limiting radius, then we may still expect some mild tidal disruption of stars in the outskirts of the GC between the Jacobi and limiting radii.
All but one (NGC 5024) of the GCs studied in this work have Jacobi radii at pericenter smaller than their King tidal radius. 
In both of these figures, we also show the Jacobi radius at apocenter to show how strong the tides are along the orbit. 
Indeed, Palomar 5 stands out as having the smallest Jacobi radius at apocenter compared to its half-light radius in Figure~\ref{fig:rtidal_vs_rhalf}. Finally, we note that five GCs have Jacobi radii at pericenter that are comparable or smaller than their half-light radius: Palomar 14, Palomar 15, ESO 452-SC11, IC 1257, and Palomar 13.

\begin{deluxetable}{ccc}
\tablehead{
\colhead{Name} & \colhead{$r_{J(peri)}$} & \colhead{$r_{J(apo)}$}\\
\colhead{} & \colhead{(pc)} & \colhead{(pc)}
}
\startdata
NGC 104 & 101.48$^{+4.88}_{-5.96}$ & 134.88$^{+2.24}_{-1.18}$ \\
NGC 288 & 20.09$^{+5.88}_{-3.50}$ & 107.32$^{+2.61}_{-1.43}$\\
\nodata & \nodata & \nodata \\
\enddata
\caption{A compilation of the Jacobi (tidal) radii used in Figures~\ref{fig:rtidal_vs_rhalf} and~\ref{fig:rtidal_vs_rKing}. 
The full Table will be available in machine-readable format upon publication; a portion is shown here for guidance regarding its form and content.}
\label{tab:jacobi}
\end{deluxetable}

\section{Summary \& Conclusions} \label{sec:conclusion}
We summarize this work as follows:
\begin{itemize}
    \item We conduct an observational search for extra-tidal envelopes and tails around 19 GCs in the DELVE DR2 footprint. We update the literature by reporting the detection of an extra-tidal envelope around NGC 5897, and report a tentative detection around NGC 7492.
    \item We run dynamical simulations to help interpret these extra-tidal envelopes, test their connection to underlying tidal tails, and forecast how deeper photometry may augment these studies. 
    Specifically, for NGC 5897, we examine whether the extra-tidal envelope plausibly indicates the presence of fainter tidal tails.
    We repeat this exercise for Pal 5, a system that has known, prominent tidal tails, as a validity check that our simulations produce tails that become more prominent at deeper magnitudes.
    We also simulate NGC 5634, a system with no known extra-tidal features, to assess whether a non-detection of extra-tidal features at DELVE DR2 depth may still be consistent with fainter tails. 
    \item We find that the observed extra-tidal envelope around NGC 5897 is consistent with the presence of fainter tidal tails. Based on our simulation, such tails would barely be detectable at the DELVE DR2 depth, even with no foreground contamination.
    An underlying stream may principally be present at higher stellar density at the full LSST depth, but the presence of Milky Way foreground could obscure this feature.
    Additional information (e.g., orbit) and altered search parameters (e.g., smoothing kernel) may aid in its detection. 
    \item Broadly, our modeling suggests that a lack of extra-tidal features at the DELVE DR2 depth does not preclude a GC from having faint tidal tails.
    Specifically, for NGC 5634, we find this to be the case.
    We also find that overdensities along the debris track may become visible at the full LSST depth.
    For NGC 5634, as with NGC 5897, the presence of Milky Way foreground meaningfully obscures these tidal tails, highlighting the potential importance of methods to separate foreground from GC/stream members in upcoming surveys.
\end{itemize}

Our results collectively motivate searches in the outskirts of GCs for extra-tidal envelopes and tidal tails with deeper photometric data, since our detections and simulations suggest that deeper data ought to contain stars that belong to faint features around these objects. 
Increasing the sample of known tidal tails around GCs will be useful for a range of science cases, spanning star cluster evolution to tracing the dark matter potential of the Milky Way. 
We demonstrate the existence of a new extra-tidal envelope around NGC 5897, a tentative envelope around NGC 7492, and our subsequent modeling motivated by these results suggests that the deep Rubin/LSST photometric depth has the potential to increase the fidelity and known number of these features. 
We highlight that additional methods to clean foreground contamination, such as proper motions and photometric metallicities (as discussed in the last paragraph of Section~\ref{sec:lsstprospects}), will likely be valuable to fully leverage deeper photometric data to prominently detect these faint features.

\begin{acknowledgments}

A.C. is supported by the Brinson Foundation through a Brinson Prize Fellowship grant.
KT thanks the Milky Way Stars group at Columbia University and the Nearby Universe group at the Center of Computational Astrophysics for useful discussions.

The DECam Local Volume Exploration Survey (DELVE; NOAO Proposal ID 2019A-0305, PI: Drlica-Wagner) is partially supported by Fermilab LDRD project L2019-011 and the NASA Fermi Guest Investigator Program Cycle 9 No. 91201.
This material is based upon work supported by the National Science Foundation under Grant No. AST-2108168, AST-2108169, AST-2307126, and AST-2407526.

This project used data obtained with the Dark Energy Camera (DECam), which was constructed by the Dark Energy Survey (DES) collaboration. Funding for the DES Projects has been provided by the U.S. Department of Energy, the U.S. National Science Foundation, the Ministry of Science and Education of Spain, the Science and Technology Facilities Council of the United Kingdom, the Higher Education Funding Council for England, the National Center for Supercomputing Applications at the University of Illinois at Urbana–Champaign, the Kavli Institute of Cosmological Physics at the University of Chicago, the Center for Cosmology and Astro-Particle Physics at the Ohio State University, the Mitchell Institute for Fundamental Physics and Astronomy at Texas A\&M University, Financiadora de Estudos e Projetos, Fundação Carlos Chagas Filho de Amparo à Pesquisa do Estado do Rio de Janeiro, Conselho Nacional de Desenvolvimento Científico e Tecnológico and the Ministério da Ciência, Tecnologia e Inovação, the Deutsche Forschungsgemeinschaft and the Collaborating Institutions in the Dark Energy Survey.

The Collaborating Institutions are Argonne National Laboratory, the University of California at Santa Cruz, the University of Cambridge, Centro de Investigaciones Enérgeticas, Medioambientales y Tecnológicas–Madrid, the University of Chicago, University College London, the DES-Brazil Consortium, the University of Edinburgh, the Eidgenössische Technische Hochschule (ETH) Zürich, Fermi National Accelerator Laboratory, the University of Illinois at Urbana-Champaign, the Institut de Ciències de l'Espai (IEEC/CSIC), the Institut de Física d'Altes Energies, Lawrence Berkeley National Laboratory, the Ludwig-Maximilians Universität München and the associated Excellence Cluster Universe, the University of Michigan, the National Optical Astronomy Observatory, the University of Nottingham, the Ohio State University, the OzDES Membership Consortium, the University of Pennsylvania, the University of Portsmouth, SLAC National Accelerator Laboratory, Stanford University, the University of Sussex, and Texas A\&M University.

Based in part on observations at Cerro Tololo Inter-American Observatory, National Optical Astronomy Observatory, which is operated by the Association of Universities for Research in Astronomy (AURA) under a cooperative agreement with the National Science Foundation.

Database access and other data services are hosted by the Astro Data Lab at the Community Science and Data Center (CSDC) of the National Science Foundation's National Optical Infrared Astronomy Research Laboratory, operated by the Association of Universities for Research in Astronomy (AURA) under a cooperative agreement with the National Science Foundation.

Fermilab is managed by FermiForward Discovery Group, LLC under Contract No. 89243024CSC000002 with the U.S. Department of Energy, Office of Science, Office of High Energy Physics. The United States Government retains and the publisher, by accepting the article for publication, acknowledges that the United States Government retains a non-exclusive, paid-up, irrevocable, world-wide license to publish or reproduce the published form of this manuscript, or allow others to do so, for United States Government purposes

This research award is partially funded by a generous gift of Charles Simonyi to the NSF Division of Astronomical Sciences. The award is made in recognition of significant contributions to Rubin Observatory’s Legacy Survey of Space and Time.

\end{acknowledgments}

\bibliography{main}{}
\bibliographystyle{aasjournal}

\end{document}